\documentclass[11pt,a4paper]{article}
\pdfoutput=1
\usepackage{jheppub}
\usepackage[utf8]{inputenc}
\usepackage{epstopdf}
\usepackage{graphicx}
\usepackage{epsfig}
\usepackage{dcolumn}  
\usepackage{bm}    
\usepackage{caption}
\usepackage{subcaption}
\usepackage{amssymb} 
\usepackage{amsmath,bm}
\usepackage{amsfonts}  
\usepackage{amsmath}  
\usepackage{slashed}  
\usepackage{enumitem}
\usepackage[mathscr]{euscript}
\usepackage{tabu}
\usepackage{epsfig}
\makeatletter
\g@addto@macro\bfseries{\boldmath}
\makeatother

\def\l1{{{1-loop}}}

\def\n1{\Bigg|_{n=1}}

\def\n{{(n)}}

\usepackage[T1]{fontenc} 
\usepackage{tikz}
\usepackage{amsmath,amssymb}
\usepackage{relsize}
\usepackage{latexsym}
\usepackage{leftidx}
\usepackage{diagbox}
\usepackage[T1]{fontenc}
\usepackage{array}
\usepackage{makecell}
\usepackage{csquotes}
\usepackage{tikz}
\usepackage{enumitem}
\usepackage{setspace}
\usepackage{multirow}
\usepackage{amsmath,amssymb}
\usepackage{relsize}
\usepackage{latexsym}
\usepackage{leftidx}
\usepackage{csquotes}
\usepackage{tikz}
\usepackage{enumitem}
\usetikzlibrary{decorations.markings}
\usetikzlibrary{decorations.pathmorphing}
\usetikzlibrary{decorations.markings}
\usetikzlibrary{decorations.pathmorphing}
\usepackage{pifont}

\usepackage{bookmark}
  \title{\textbf{\textsf{Partition functions and entanglement entropy: Weyl graviton and conformal higher spin fields}}}

\author{ Jyotirmoy Mukherjee}
\affiliation{\vspace{.1cm} Centre for High Energy Physics, \\ Indian Institute of Science,\\
C. V. Raman Avenue, Bangalore 560012, India.}
\emailAdd{ jyotirmoym@iisc.ac.in}
\abstract{We establish the relation of partition functions of conformal higher spin fields on Weyl equivalent spaces in $d=4$ dimension. We express the partition function of Weyl graviton and conformal higher spin fields as an integral over characters on  $S^1\times AdS_3$, $S^4$, and  $AdS_4$. We observe that the partition function of conformal higher spins on hyperbolic cylinders differs from the partition function on $S^4$ by the `edge' contribution. The logarithmic coefficient obtained from the character integral of the partition function of conformal higher spins on $AdS_4$ is the half of that obtained from the partition function on $S^4$.  We evaluate the entanglement entropy and the conformal dimension of the twist operator from the partition function on the hyperbolic cylinder. The conformal dimension of the co-dimension two twist operator enables us to find a linear relation between Hofman-Maldacena variables which we use to show the non-unitarity of the theory. We observe that the spectrum of the quasinormal modes of conformal higher spins obtained from the bulk character contains additional distinct states compared to the spectrum of unitary massless higher spin fields. }
\begin{document}

\maketitle
The one-loop partition functions of free conformal fields on spheres, hyperbolic cylinders, and anti-de Sitter spaces have helped us to understand the horizon entropy of the theories of gravity coupled to matter. This was initiated in \cite{Gibbons:1976ue} and more recent developments can be found \cite{Anninos:2021ene, Anninos:2021ihe, Anninos:2021ydw,Sun:2020ame,Pethybridge:2021rwf,Grewal:2021bsu}.
It is well known that the spheres, hyperbolic cylinders, and anti-de Sitter spaces are related by Weyl transformation. Since the spaces are conformally related, one can ask how the partition functions of conformal fields are related to these backgrounds. A similar kind of question was asked before in \cite{Klebanov:2011uf} where it was shown that the partition functions of conformal scalars and fermions are identical on $S^3$ and $S^1\times AdS_2$. Recently in \cite{Anninos:2020hfj, David:2021wrw}, it was shown that partition functions of scalars, fermions, integer higher spin fields, and anti-symmetric $p$-forms can be expressed as integral over bulk and edge characters on a Euclidean patch of de Sitter and  anti-de Sitter spaces. In \cite{David:2021wrw} the partition function of conformal $p$-forms was compared on the sphere and hyperbolic cylinders and it was shown that hyperbolic cylinders capture only the bulk partition function of the sphere. Similarly in \cite{Mukherjee:2021alj}, it was observed that the partition function of a higher derivative conformal vector field on $S^1\times AdS_5$ agrees with the bulk partition function of $S^6$ and misses out the edge part. The edge partition function of the higher derivative conformal vector field on $S^6$ was identified with the higher derivative conformal scalar on $S^4$. The similar pattern was also found in \cite{David:2021wrw} for conformal $p$-forms.

In this paper, we evaluate the partition function of Weyl graviton and free conformal higher spins on conformally related spaces in $d=4$ dimension and find relationships among them. We express the partition function of Weyl graviton and conformal higher spin fields in terms of bulk and edge characters on $S^4$. We also evaluate the partition function of the same on $S^1\times AdS_3$ and find that it can be written as integral over only the bulk character of the partition function and misses out the edge mode.
\begin{align}
    -\log\mathcal{Z}_{s,\rm{Weyl}}^{(S^4)}&=\int_0^{\infty}\frac{dt}{2t}\frac{1+x}{1-x}\left(\chi^{b,\rm{dS}}_{s,\rm{Weyl}}-\chi^{e,\rm{dS}}_{s,\rm{Weyl}}\right); \qquad x=e^{-t}.
\end{align}
Here $\chi^{b,\rm{dS}}_{s,\rm{Weyl}}$ and $\chi^{e,\rm{dS}}_{s,\rm{Weyl}}$ are the bulk and edge characters of conformal spin-$s$ filed on $S^4$ respectively. But the partition function of conformal spin-$s$ field on $S^1\times AdS_3$ consists only the bulk part.
\begin{align}
   -\log\mathcal{Z}_{s,\rm{Weyl}}[S^1\times AdS_3]& =\frac{\log R}{2\pi i}\int_C\frac{dt}{2t}\frac{1+x}{1-x}\chi^{b,\rm{dS}}_{s,\rm{Weyl}}    ; \qquad x=e^{-t},
\end{align}
where $C$ is the integration contour on the $t$-plane and $R$ is the ratio of the radial cutoff on $AdS$ to the radius of $AdS$. 

The partition function of conformal fields on hyperbolic cylinders is useful to evaluate entanglement entropy across a spherical surface. Therefore we obtain the free energy of the conformal spin-$s$ field on the hyperbolic cylinder with the radius along the $S^1$ direction is $2\pi q$. One can think of it as a thermal ensemble with the inverse temperature is related to the radius of the circle. The  R\'{e}nyi entropy and entanglement entropy across a spherical surface of conformal higher spins can be obtained from the thermal free energy on hyperbolic cylinders \cite{Hung:2014npa}
\begin{align}
    \begin{split}
         S_{s,\rm{Weyl}}(q)&=\frac{-\mathcal{F}_{q,s}+q\mathcal{F}_{q=1,s}}{1-q}, \qquad\qquad
          S^{\rm{EE}}_{s,\rm{Weyl}}=\lim_{q\rightarrow 1} S_{s,\rm{Weyl}}(q).
  \end{split}
\end{align}
Here $\mathcal{F}_{q,s}$ is the free energy of conformal higher spins on the hyperbolic cylinder. 
We compute R\'{e}nyi entropy and entanglement entropy of Weyl graviton and conformal higher spins in $d=4$ dimension using the character integral representation of free energy on the hyperbolic cylinder. The universal contribution of the entanglement entropy of Weyl graviton is given by
\begin{align}
 S_{2,\rm{Weyl}}(q)=-\frac{(q+1) \left(71 q^2+1\right)}{60 q^3}\log R,\qquad  S^{\rm{EE}}_{2,\rm{Weyl}}=-\frac{12}{5}\log R.
\end{align}
The expression of the  R\'{e}nyi entropy and entanglement entropy of conformal higher spins are presented in section \eqref{spinee}. 
Similarly, we express the partition function of conformal higher spin fields in terms of integral over characters on $AdS_4$ and show that the logarithmic divergent  piece of the partition function is the half of that obtained from the partition function on $S^4$.
We evaluate the conformal dimension of the twist operator which determines the energy density  across the entangling surface in the presence of a defect.
\begin{align}\label{regv}
    h_{q,s}=\frac{q}{3\rm{Vol}(AdS_3)}(\partial_q \mathcal{F}_{q,s}-\partial_q \mathcal{F}_{q,s}|_{q=1}), \qquad  \rm{Vol}(AdS_3)=2\pi\log R.
\end{align}
Here we have used the regularised volume of $AdS_3$ given in \cite{Hung:2014npa}.

Conformal higher spin theories admit higher derivative kinetic terms in the action and therefore have negative residue in the propagator \cite{Donoghue:2017fvm}. The negative residue implies a negative norm state and therefore one would think the conformal higher spin theories are non-unitary. We use the causality bound of the Hofman-Maldacena variables \cite{Camanho:2009vw} to show the non-unitarity property of the conformal higher spin fields. The positivity of the energy flux or average null energy condition puts bounds on the parameters $t_2$ and $t_4$. This bound represents a triangular region in the $t_2-t_4$-plane and the unitary theories represent points inside this region. The first and second derivatives of the conformal dimension of the twist operator can be related to the two and three-point functions of the stress tensors \cite{Huang:2014pfa} which can be expressed in terms of the three parameters $a$, $b$, and $c$ \cite{Osborn:1993cr}. The ratio of these parameters is related to the Hofman-Maldacena variables $t_2$ and $t_4$. From the explicit computation of the first and second derivative of the conformal dimension of the twist operator, we find a linear relation of the Hofman-Maldacena variables which represents a straight line in $t_2-t_4$-plane. We find that this straight line never intersects the region of unitarity in the plane and therefore the theory is non-unitary.

We extract the quasinormal spectrum of the static $dS$ patch from the bulk character of conformal higher spin fields. We compare it with the quasinormal spectrum of unitary massless higher spin fields on $S^4$ \cite{Anninos:2020hfj} and observe that the spectrum of quasinormal modes of conformal higher spin fields has $(s-1)$ number of extra modes compared to the spin-$s$ unitary massless fields. In \cite{Sun:2020ame} it was shown that the character constructed from the quasinormal spectrum agrees with the Harish-Chandra character of the unitary massless higher spin fields. Therefore we conclude that the quasinormal spectrum of conformal higher spins consists of extra distinct states compared to the massless unitary higher spin fields. 
\section{Weyl graviton in $d=4$ dimension}
We study the one-loop partition function of Weyl graviton on conformally related spaces in $d=4$ dimension. We show that the partition function of Weyl graviton on $S^4$, $S^1\times AdS_3$, and $AdS_4$ can be expressed as integral over characters.  We observe that the logarithmic coefficient obtained from the character integral representation of the partition function on $S^4$ is twice  that on $AdS_4$. This is also observed in the conformal scalar case. However, the hyperbolic cylinder contains only the bulk character of $S^4$ and misses out the edge character of the partition function. A similar property was also observed in the Maxwell field in $d=4$ dimension \cite{David:2020mls}. The method we develop here will be helpful to compare the partition functions of conformal higher spin fields on Weyl equivalent spaces by comparing only the integrands.
\subsection{Weyl graviton on $S^4$}
We now explain the techniques to express the partition function of Weyl graviton in terms of bulk and edge characters on $S^4$. We start with the action of Weyl graviton in $d=4$ dimension, decompose the field in the transverse and longitudinal modes, and fix the gauge to obtain the partition function. Finally, the partition function involves Laplacians of the transverse spin fields on $S^4$ which we express as the integral over transverse characters.
After dropping the total derivative term, the action of Weyl graviton in $d=4$ dimension becomes \cite{Beccaria:2014jxa}
\begin{align}\label{weylgrav}
    S&=\frac{1}{2}\int \sqrt{g}d^4x C_{\mu\nu\rho\lambda}C^{\mu\nu\rho\lambda}\nonumber\\
    &=\int \sqrt{g}d^4x \left(R_{\mu\nu}R^{\mu\nu}-\frac{R^2}{3}\right).
\end{align}
Here $C_{\mu\nu\rho\lambda}$ is the Weyl tensor and $R_{\mu\nu}$ is the Riemann tensor in $d=4$ dimension. One obtains the one-loop partition function by expanding the action to quadratic order in 
fluctuations in the background of  a conformally flat space \cite{Beccaria:2014jxa}
\begin{align}\label{eq}
    \mathcal{L}^{(2)}&=\frac{1}{4}\nabla^2h_{\mu\nu}\nabla^2h^{\mu\nu}-R^{\mu}_{\rho}h_{\mu\nu}\nabla^2h^{\nu\rho}+\frac{1}{2}R^{\mu\nu}h_{\alpha\beta}\nabla_{\mu}\nabla_{\nu}h^{\alpha\beta}-\frac{3}{2}R_{\rho\sigma}R^{\sigma\mu}h_{\mu\nu}h^{\nu\rho}\nonumber\\
    &+\frac{1}{2}R^{\nu\rho}R^{\sigma\mu}h_{\mu\nu}h_{\rho\sigma}+\frac{1}{6}(h_{\mu\nu}R^{\mu\nu})^2+\frac{1}{4}R_{\mu\nu}R^{\mu\nu}h_{\alpha\beta}h^{\alpha\beta}\nonumber\\
    &+\frac{1}{2}RR^{\mu}_{\rho}h_{\mu\nu}h^{\nu\rho}-\frac{R^2}{9}h_{\mu\nu}h^{\mu\nu}.
\end{align}
When the conformally flat space is the Einstein space, one substitutes $R_{\mu\nu}=\frac{1}{4}g_{\mu\nu}R$ and obtains
\begin{align}\label{ees}
    \mathcal{L}^{(2e)}&=\frac{1}{4}\nabla^2h_{\mu\nu}\nabla^2h^{\mu\nu}-\frac{1}{8}Rh_{\mu\nu}\nabla^2h^{\mu\nu}+\frac{1}{72}R^2h_{\mu\nu}h^{\mu\nu}\nonumber\\
    &=h_{\mu\nu}\mathcal{O}_2h^{\mu\nu},
\end{align}
where $\mathcal{O}_2$ is the fourth-order operator which can be factorized in the following way
\begin{align}\label{4op}
    \mathcal{O}_2&=\left(\Delta_2+\frac{R}{3}\right)\left(\Delta_2+\frac{R}{6}\right).
\end{align}
Here $\Delta_2$ is the spin-$2$ Laplacian acting on the transverse-traceless field. After the gauge fixing procedure and including the change in measure shown in \eqref{gaugefixs}, the partition function of Weyl-graviton on $S^4$ takes the following form \cite{Beccaria:2017lcz}
\begin{align}\label{partwes4}
   \mathcal{Z}_{2,\rm{Weyl}}^{(S^4)}&=\left(\frac{\det\Delta_{(1)}^{\perp}(-3)\det\Delta_{(0)}(-4)}{\det\Delta_{(2)}^{\perp}(4)\det\Delta_{(2)}^{\perp}(2)}\right)^{\frac{1}{2}}.
\end{align}
Here $\Delta_{(s)\perp}(M^2)$ refers to the Laplacian of transverse traceless 
spin $s$ field on $S^{4}$ with a mass term $M^2$ coming from the curvature coupling.\\
To evaluate the  logarithm of the determinant we need the degeneracy and the spectrum of the transverse spin-$s$ Laplacian on the sphere.
The eigen-value and the degeneracy on $S^4$ is given by \cite{Camporesi:1994ga}
\begin{align}\label{egdeg}
    \begin{split}
    \lambda_{n,4}^{(s)}&=n(n+3)-s,\\
        g_{n,4}^{(s)}&=\frac{1}{6} (2 n+3) (2 s+1) (n-s+1) (n+s+2).\\
    \end{split}
\end{align}
Using the eigen-value and the degeneracy,
 the $\log\det\Delta_{(s)\perp}(M_s^2)$ can be written as
\begin{align}\label{spinspartition}
- \frac{1}{2}\log\det\Delta_{(s)\perp}(M_s^2)
  & =\int_0^\infty \frac{d\tau}{2\tau}\bigg[  \left(\sum_{n =s}^\infty
    g_{n }^{(s)} (  e^{ - \tau ( \lambda_{n}^{(s)}+M_s^2)}  - e^{-\tau} )\right)
  \bigg].
\end{align}
 We have used the following identity to express the $\log\det\Delta_{(s)\perp}(M_s^2)$
 \begin{equation}\label{logiden}
-\log y  = \int_0^\infty \frac{d\tau}{\tau} ( e^{-y \tau} - e^{-\tau} ) .
\end{equation}
We evaluate the second term of \eqref{spinspartition}  by performing a sum over the degeneracy $g_{n,d}^{(s)}$ on $S^d$ at 
sufficiently negative $d$  and analytically continue it to the positive $d$ \cite{David:2021wrw}. 
Note that the large $n$ behaviour of the degeneracy is give by 
\begin{align}
    g_{n,d }^{(s) }&=\frac{(d+2 n-1) (d+2 s-3) (n-s+1) (d+n-3)! (d+s-4)! (d+n+s-2)}{((d-1)! (n+1)!) ((d-3)! s!)}\nonumber\\
    &=\left(\frac{1}{n}\right)^{-d}\left(\frac{2 (d+2 s-3) (d+s-4)!}{n (d-3)! (d-1)! s!}+\frac{\left(d^2-2 d+1\right) (d+2 s-3) (d+s-4)!}{n^2 (d-3)! (d-1)! s!}+\mathcal{O}\left(\frac{1}{n^3}\right)\right)
\end{align}
Therefore one observes that the sum $\sum_{n=s}^{\infty}  g_{n }^{(s)}$ converges in the large negative value of $d\leq -1$ .  We evaluate the sum over degenracy of transverse spin-$s$ field and observe that 
\begin{align} \label{sumrelten}
    \sum_{n=s}^{\infty}  g_{n,d }^{(s)}
    &=\frac{(d+2 s-4) (d+2 s-3) (d+2 s-2) \Gamma (d+s-3) \Gamma (d+s-2)}{s! \Gamma (d-1) \Gamma (d+1) \Gamma (s)}\nonumber\\
    &=-\sum_{n=-1}^{s-1}  g_{n,d}^{(s)}.
\end{align}
Therefore we can  extend the sum from $n=-1$ to $n=\infty$.
Due to this relation  we can re-write \eqref{spinspartition} as
\begin{align}\label{repstep}
- \frac{1}{2}\log\det\Delta_{(s)}^{\perp}(M_s^2)
  & =\int_0^\infty \frac{d\tau}{2\tau}e^{-\frac{\epsilon^2}{4\tau}} \bigg[ \left(\sum_{n =-1}^\infty
    g_{n }^{(s)} (  e^{ - \tau ( \lambda_{n}^{(s)}+M_s^2)}  \right) \bigg].
\end{align}
At this point, we introduce the factor  $e^{-\frac{\epsilon^2}{4\tau} } $  which will help us to understand the branch cut in the integration plane. We have not introduced this to act as a regulator because  the starting integral  \eqref{spinspartition} is already convergent after dropping the second term by using the dimension regularisation \cite{David:2021wrw}. 
 We now use the Hubbard-Stratonovich trick to perform the sum over the eigen modes of the transverse traceless spin-$s$ field on $S^4$
    \begin{equation} \label{contint}
   \sum_{n=-1}^{\infty}g_{n}^{(s)} e^{-\tau(n+\frac{3}{2})^2}
   =\int_{C} \frac{du}{\sqrt{4\pi\tau}}e^{-\frac{u^2}{4\tau}}f_s(u).
\end{equation}
Here the contour $C$ runs from $-\infty $ to $\infty$ which is lifted slightly above the real line and $f_s(u)$ is given by
 \begin{align}\label{spinsum}
        f^{(s)}(u)&=\sum_{n=-1}^{\infty}g^{(s)}_n e^{i u(n+\frac{3}{2}}).
    \end{align}

     We  perform the  integral over $\tau$ in \eqref{repstep} which results in 
     \begin{align}
       - \frac{1}{2}\log\det\Delta_{(s)}^{\perp}(M_s^2) &=   \int_{C}\frac{du}{ 2\sqrt{u^2+\epsilon^2}}\left(e^{-\nu_s\sqrt{u^2+\epsilon^2}}f_s(u)\right). 
     \end{align}
       with $\nu^2_s=-(-M_s^2+s+\frac{9}{4})$ .\\
     Now we deform the contour 
$C$ to the contour  $C'$ which folds around the branch cut on the imaginary axis 
originating at $u = i \epsilon$ on the  $u$-plane. This is shown in figure [\ref{fig1}].
Substituting $u = i t$ we obtain
    \begin{eqnarray} \label{spin s simplified}
     - \frac{1}{2}\log\det\Delta_{(s)}^{\perp}(M_s^2) & = \int_{\epsilon}^{\infty}\frac{dt}{2\sqrt{t^2-\epsilon^2}}\bigg[\left(e^{i\nu_s\sqrt{t^2-\epsilon^2}}+e^{-i\nu_s\sqrt{t^2-\epsilon^2}}\right)f^{(s)} (i t)
      \bigg].
    \end{eqnarray}
  
\begin{figure}[h]
\centering
\begin{tikzpicture}[thick,scale=0.85]
\filldraw[magenta] 
                (0,0.5) circle[radius=3pt]
                (0,-0.5) circle[radius=3pt];
                \filldraw[orange] 
                (0.1,0) circle[radius=2pt]
                (0.5,0) circle[radius=2pt]
              (1,0) circle[radius=2pt]  
              (1.5,0) circle[radius=2pt]
              (2,0) circle[radius=2pt]
               (2.5,0) circle[radius=2pt]
                (3,0) circle[radius=2pt]
               (3.5,0) circle[radius=2pt]
                (4,0) circle[radius=2pt]
                (-0.1,0) circle[radius=2pt]
                (-0.5,0) circle[radius=2pt]
              (-1,0) circle[radius=2pt]  
              (-1.5,0) circle[radius=2pt]
              (-2,0) circle[radius=2pt]
               (-2.5,0) circle[radius=2pt]
                (-3,0) circle[radius=2pt]
               (-3.5,0) circle[radius=2pt]
                (-4,0) circle[radius=2pt] ;
\draw [decorate,decoration=snake] (0,-0.5) -- (0,-4);
\draw [decorate,decoration=snake] (0,0.5) -- (0,4);
\draw [postaction={decorate,decoration={markings , 
mark=at position 0.55 with {\arrow[black,line width=0.5mm]{<};}}}](0.2,1) arc[start angle=0, end angle=-180, radius=0.2cm];
\draw
[
postaction={decorate,decoration={markings , 
mark=at position 0.20 with {\arrow[red,line width=0.5mm]{>};}}}
][blue, thick] (-4,0.2)--(0,0.2);
\draw[blue, thick] (0,0.2)--(4,0.2);
\draw (0.4,1) node{$\mathbf{\epsilon}$};
\draw (0.4,-1) node{$\mathbf{-\epsilon}$};
\draw[gray, thick] (0,0) -- (0,4);
\draw[gray, thick] (0,0) -- (0,-4);
\draw[gray, thick] (0.2,4) -- (0.2,1);
\draw[gray, thick] (-0.2,1) -- (-0.2,4);
\draw[gray, thick] (0,0) -- (4,0);
\draw[gray, thick] (0,0) -- (-4,0);
\draw[gray, thick] (2,3) -- (2,3.4);
\draw[gray, thick] (2,3) -- (2.5,3);
\draw (2.3,3.3) node{$\mathbf{u}$};
\draw [blue,thick](-2,0.5) node{$\mathbf{C}$};
\draw [gray,thick](0.5,3) node{$\mathbf{C'}$};
\end{tikzpicture}
\caption{Integration contour in $u$-plane} \label{fig1}
\qquad
\centering
\begin{tikzpicture}[thick,scale=0.85]
\filldraw[magenta] 
                (0.5,0) circle[radius=3pt]
                (-0.5,0) circle[radius=3pt];
                \filldraw[orange] 
               ( 0,0.2) circle[radius=2pt]
                (0,0.5) circle[radius=2pt]
              (0,1) circle[radius=2pt]  
              (0,1.5) circle[radius=2pt] 
              (0,2) circle[radius=2pt] 
              (0,2.5) circle[radius=2pt]
                (0,3) circle[radius=2pt]
               (0,3.5) circle[radius=2pt]
                (0,4) circle[radius=2pt]
                (0,-0.1) circle[radius=2pt]
                (0,-0.5) circle[radius=2pt]
              (0,-1) circle[radius=2pt]  
              (0,-1.5) circle[radius=2pt]
              (0,-2) circle[radius=2pt]
               (0,-2.5) circle[radius=2pt]
                (0,-3) circle[radius=2pt]
               (0,-3.5) circle[radius=2pt]
                (0,-4) circle[radius=2pt] ;
\draw [decorate,decoration=snake] (-0.5,0) -- (-4,0);
\draw [decorate,decoration=snake] (0.5,0) -- (4,0);
\draw
[
postaction={decorate,decoration={markings , 
mark=at position 0.20 with {\arrow[red,line width=0.5mm]{>};}}}
][blue, thick] (0.5,0.2)--(4,0.2);
\draw (1,0.4) node{$\mathbf{\epsilon}$};
\draw (-1,0.4) node{$\mathbf{-\epsilon}$};
\draw[gray, thick] (0,0) -- (0,4);
\draw[gray, thick] (0,0) -- (0,-4);
\draw[gray, thick] (0,0) -- (4,0);
\draw[gray, thick] (-4,0) -- (0,0);
\draw[gray, thick] (2,3) -- (2,3.4);
\draw[gray, thick] (2,3) -- (2.5,3);
\draw (2.3,3.3) node{$\mathbf{t}$};
\end{tikzpicture}
\caption{Integration contour in $t$-plane} \label{fig2}
\end{figure}

     Substituting \eqref{spinsum} in  \eqref{spin s simplified} and taking the limit $\epsilon\rightarrow 0$ we obtain
 \begin{align} \label{parttensor}
   - \frac{1}{2}\log\det\Delta_{(s)}^{\perp}(M_s^2)  & =\int_0^{\infty}\frac{dt}{2t}\left(e^{i\nu_s t}+e^{-i\nu_s t}\right)f^{(s)} (i t).
 \end{align}
 Using the relation in \eqref{parttensor} we express the logarithm of determinants of all transverse traceless fields which appear in the expression of the partition function of Weyl graviton on $S^4$
 \begin{align}
     \begin{split}
        - \frac{1}{2}\log\det\Delta_{(2)}^{\perp}(4)  & =\int_0^{\infty}\frac{dt}{2t}\frac{1+x}{1-x}\left(\frac{5 \left(x^2+x\right)}{(1-x)^3}-\frac{5 (x+1)}{1-x}\right),\\
          - \frac{1}{2}\log\det\Delta_{(2)}^{\perp}(2)  &=\int_0^{\infty}\frac{dt}{2t}\frac{1+x}{1-x}\left(\frac{5 \left(x^3+1\right)}{(1-x)^3}-\frac{5 \left(x^2+\frac{1}{x}\right)}{1-x}\right),\\
         - \frac{1}{2}\log\det\Delta_{(1)}^{\perp}(-3)  &=\int_0^{\infty}\frac{dt}{2t}\frac{1+x}{1-x}\left(\frac{3 \left(x^4+\frac{1}{x}\right)}{(1-x)^3}-\frac{x^3+\frac{1}{x^2}}{1-x}\right),\\
        - \frac{1}{2}\log\det\Delta_{(0)}^{\perp}(-4)  & =\int_0^{\infty}\frac{dt}{2t}\frac{1+x}{1-x}\frac{x^4+\frac{1}{x}}{(1-x)^3}  .
     \end{split}
 \end{align}
 Here $x=e^{-t}$. Note that $\log\det\Delta_{(s)}^{\perp}(M_s^2)$ has been expressed in terms of integral over bulk and edge characters. The bulk and the edge characters of the transverse traceless spin-$s$ field fit in the following form on  $S^4$.  This is consistent with the characters of massive transverse spin-$s$ field on $S^4$ \cite{Anninos:2020hfj} but here mass terms arise from the curvature coupling. 
 \begin{eqnarray}
    \chi^{b,\rm{dS}}_{s,\perp}=5\frac{x^{\frac{3}{2}+i\nu_s}+x^{\frac{3}{2}-i\nu_s}}{(1-x)^3},\quad  \chi^{e,\rm{dS}}_{s,\perp}=\frac{x^{\frac{1}{2}+i\nu_s}+x^{\frac{1}{2}-i\nu_s}}{(1-x)}.
 \end{eqnarray}
Combining all the determinants of the transverse traceless fields appearing in the expression of  the partition function of the Weyl graviton, we obtain
 \begin{align}\label{grav}
  -\log   \mathcal{Z}_{2,\rm{Weyl}}^{(S^4)}&=\int_0^{\infty}\frac{dt}{2t}\frac{1+x}{1-x}\bigg[\left(\frac{5 \left(x^2+x\right)}{(1-x)^3}+\frac{5 \left(x^3+1\right)}{(1-x)^3}-\frac{3 \left(x^4+\frac{1}{x}\right)}{(1-x)^3}-\frac{x^4+\frac{1}{x}}{(1-x)^3} \right)\nonumber\\
  &-\left(\frac{5 (x+1)}{1-x}+\frac{5 \left(x^2+\frac{1}{x}\right)}{1-x}-\frac{x^3+\frac{1}{x^2}}{1-x}\right)\bigg].
 \end{align}
 Therefore we get the full `naive' bulk and the edge characters of Weyl gravion on $S^4$ 
 \begin{align}
    \chi^{b,\rm{dS}}_{2,\rm{Weyl}}&=\left(\frac{5 \left(x^2+x\right)}{(1-x)^3}+\frac{5 \left(x^3+1\right)}{(1-x)^3}-\frac{3 \left(x^4+\frac{1}{x}\right)}{(1-x)^3}-\frac{x^4+\frac{1}{x}}{(1-x)^3} \right),\\
     \chi^{e,\rm{dS}}_{2,\rm{Weyl}}&=\left(\frac{5 (x+1)}{1-x}+\frac{5 \left(x^2+\frac{1}{x}\right)}{1-x}-\frac{x^3+\frac{1}{x^2}}{1-x}\right).
 \end{align}

 So the partition function of the Weyl graviton on $S^4$ becomes the integral over the `naive' bulk and the `naive' edge character. This is going to be useful to compare the partition function of Weyl graviton on the spaces which are conformally related because we just compare the integrands. We also evaluate the logarithmic divergent piece of the partition function by expanding the integrand \eqref{grav} around $t=0$ and collecting $1/t$ term in the expansion which turns out to be
$
     \alpha_{s=2}^{(4)}=-\frac{87}{5}.
$

Note that this naive character  grows as $e^{t}$ and therefore 
cannot be considered as  character of a unitary representation of $SO(1, 4)$ \cite{Anninos:2020hfj}. Therefore we replace the `naive' characters by the `flipped characters'. Consider the character $\chi = \sum_{\ell} c_{\ell} x^{\ell}  $ with terms $\ell<0$, then the 
`flipped character' is  given by 
\begin{equation} \label{flip}
[\chi]_+ =  \chi - c_0 - c_{\ell} ( x^ {\ell} + x^{ -\ell} ) .
\end{equation}
As explained in \cite{Anninos:2020hfj}, this is basically  a contour deformation 
so that the integration is done over the negative $t$ axis.  This procedure is known as the flipping of the naive character. This `flipped character' is actually the unitary irreducible representation of $SO(1,4)$. So one can in principle replace them with the `flipped characters'. But for our purpose, we compare the partition functions with the `naive' characters which will be sufficient to find the relation among them. 
 \subsection{Weyl graviton on $AdS_3\times S^1_q$}
 The action of the Weyl graviton in $d=4$ dimension is given in \eqref{weylgrav}. To obtain the partition function on the background of $AdS_3\times S^1$, we expand the action to the quadratic order in the same background. We adapt the procedure developed in \cite{Beccaria:2014jxa} to fix the gauge.
 We first split the components of $h_{\mu\nu}$ into $h_{00}$, $h_{0i}$ and $h_{ij}$, where $\{ i,j\}$ are the directions along $AdS_3$. We further decompose $h_{ij}$ into traceless and pure trace part.
\begin{eqnarray}\label{defth}
h_{ij}  &=& \bar h_{ij} + \frac{1}{3} g_{ij} h,  \\ \nonumber
\bar h_{ij}  &=& h_{ij}^\perp + \nabla_i \zeta^\perp_j + \nabla_j\zeta^\perp_i  +
 \nabla_i \nabla_j \sigma -\frac{1}{3} g_{ij} \Box\sigma ,  \\ \nonumber
 & &{\rm where} \qquad  \nabla^i h_{ij}^\perp = 0, \qquad \nabla^i \zeta^\perp_i =0.
\end{eqnarray}
In the second line, we  decompose the traceless part into transverse and longitudinal parts. The transverse $h^{\perp}_{ij}$ satisfies the transverse gauge condition given above.
We use $R_{00}=0$, $R_{0i}=0$ and $R_{ij}=\frac{1}{3}g_{ij}R$ and simplify the action. We now substitute these in \eqref{eq} and 
 the transverse traceless part of the Lagrangian takes the following form
 \begin{align}
     \mathcal{L}_{h^{\perp}_{ij}}&=-\frac{1}{4}h^{\perp ij}(\partial_0^2+\Delta^{\rm{AdS}}_2)^2h^{\perp}_{ij}+2h^{\perp}_{ij}\partial_0^2h^{\perp ij}+h^{\perp ij}\Delta^{\rm{AdS}}_2h^{\perp}_{ij}+h^{\perp ij}h^{\perp}_{ij}\nonumber\\
     &=-\frac{1}{4}h^{\perp ij}\bigg[\partial_0^2-\left(\left(-\Delta^{\rm{AdS}}_{2}-3\right)^{\frac{1}{2}}+i\right)^2\bigg]\bigg[\partial_0^2-\left(\left(-\Delta^{\rm{AdS}}_{2}-3\right)^{\frac{1}{2}}-i\right)^2\bigg]h^{\perp}_{ij}.
 \end{align}
 In the last line, we have factorized the fourth-order operator in terms of two quadratic order operators. Note that this is not the Einstein space and therefore we used the Lagrangian \eqref{eq} instead of \eqref{ees}.
After fixing the gauge and considering Faddeev-Popov determinants and change in measure, shown in appendix \eqref{appa}, the partition function becomes
 \begin{align}\label{weylpart}
     \mathcal{Z}_{2,\rm{Weyl}}[{AdS_3\times S^1_q}]&=\bigg[\partial_0^2-\left(\left(-\Delta^{\rm{AdS}}_{2}-3\right)^{\frac{1}{2}}+i\right)^2\bigg]^{-\frac{1}{2}}\bigg[\partial_0^2-\left(\left(-\Delta^{\rm{AdS}}_{2}-3\right)^{\frac{1}{2}}-i\right)^2\bigg]^{-\frac{1}{2}}\nonumber\\
    & \times \left(\partial_0^2-\Delta^{\rm{AdS}}_{1}-2\right)^{-\frac{1}{2}},
 \end{align}

 where $\Delta^{\rm{AdS}}_{2}$ is the transverse spin-2 Laplacian on $AdS_3$ with the eigen value \cite{Camporesi:1994ga}
 \begin{align}
     -\Delta^{\rm{AdS}}_{2}\psi^{(\lambda,u)}=(\lambda^2+3)\psi^{(\lambda,u)}.
 \end{align}
 $AdS$ is a non-compact space and the eigenvalue spectrum is distributed through a measure known as Plancherel measure. The expression of the Plancherel measure of transverse spin-$s$ field on $AdS_3$ is given by \cite{Camporesi:1994ga}
 \begin{align}
     d\mu_{s}^{(3)} &= \frac{{\rm Vol} (AdS_3)}{\pi^2} ( \lambda^2 + s^2) d\lambda.
 \end{align}
The conformal spin-$1$ part with a full tower of Kaluza-Klein mass comes from the integration over $\zeta^{i\perp}$ in the action. 
We use the identity \eqref{logiden} to replace the logarithm of the determinants and the partition function of Weyl graviton on $AdS_3\times S^1_q$ becomes
 \begin{align}
     -\log\mathcal{Z}_{2,\rm{Weyl}}[{AdS_3\times S^1_q}]&=\frac{1}{4} \int_0^\infty \frac{d\tau}{\tau} \sum_{n=-\infty}^\infty 
  \int_{-\infty}^{\infty} d\lambda 
  \mu_{2}^{(3)}(\lambda)\bigg[ ( e^{ -\tau( (\lambda+i)^2 + \frac{n^2}{q^2} )} - e^{-\tau} )   \nonumber\\
  &+( e^{ -\tau( (\lambda-i)^2 + \frac{n^2}{q^2} )} - e^{-\tau} )\bigg]\nonumber\\
  &+\frac{1}{4} \int_0^\infty \frac{d\tau}{\tau} \sum_{n=-\infty}^\infty 
  \int_{-\infty}^{\infty} d\lambda 
  \mu_{1}^{(3)}(\lambda)\bigg[ ( e^{ -\tau( \lambda^2 + \frac{n^2}{q^2} )} - e^{-\tau} )  \bigg]
 \end{align}
To evaluate the partition function,  the integrals over the Plancherel measure are needed to be computed. Note that, the Plancherel measure is an even function in $\lambda$. Therefore we extend the lower limit of the integral all the way to $-\infty$ to obtain
\begin{align*}
    \int_{-\infty}^{\infty}\mu_s^{(3)}(\lambda)d\lambda=\lim_{u\rightarrow 0}W_s^{(3)}(u).
\end{align*}
Here $W_s^{(d)}(u)$ is the Fourier transform of the Plancherel measure of transverse spin-$s$ field on $AdS_d$. The expression of $W_s^{(d)}$ is given by \cite{Sun:2020ame}
\begin{align}\label{ftran}
  W_s^{(d)}( u)&=  \left(\binom{d+s}{d}-\binom{d+s-2}{d}\right)\times\nonumber\\
  &\frac{(-1)^{d+1} (d-1)! (e^{-u}+1) e^{-(\frac{d}{2}+1)u}\left(s (d+s-1) (1-e^{-u})^2-d (d+1) e^{-u}\right)}{(1-e^{-u}) (1-e^{-u})^{d+1}}
\end{align}
From the expression of $W_s^{(d)}( u)$  we observe  that $\lim_{u\rightarrow 0}W_s^{(d)}(u)$ vanishes for sufficiently large negative $d$. Therefore the integral of the Plancherel measure of transverse spin-$s$ field on $AdS$ also vanishes. A similar treatment can be found for co-exact $p$-forms in \cite{David:2021wrw}.
\begin{equation}
\int_0^\infty d\lambda \mu_s^{(d)} =0 .
\end{equation}
 We proceed with the first term and obtain
\begin{align}
 -\log\mathcal{Z}_{2,\rm{Weyl}}[{AdS_3\times S^1_q}]&
  =
  \int_0^{\infty}\frac{d\tau}{4\tau}e^{-\frac{\epsilon^2}{4\tau}}\bigg[\int_{-\infty}^{\infty} 
  d\lambda 
  \mu^{(3)}_2 (\lambda) \big[ \left(  e^{ - \tau (\lambda+i)^2} + 2 \sum_{n=1}^\infty 
  e^{-\tau(\lambda^2+\frac{n^2}{q^2})  }\right)\nonumber\\
  &+\left(  e^{ - \tau (\lambda-i)^2} + 2 \sum_{n=1}^\infty 
  e^{-\tau(\lambda^2+\frac{n^2}{q^2})  }\right)\big]+\int_{-\infty}^{\infty} 
  d\lambda 
  \mu^{(3)}_1 (\lambda)  \left(  e^{ - \tau \lambda^2} + 2 \sum_{n=1}^\infty 
  e^{-\tau\lambda^2+\frac{n^2}{q^2})  }\right)\bigg]
\end{align}
We perform the  integral over $\lambda$ using the Hubbard-Stratonovich trick which linearises the sum over Kaluza-Klein modes along $S^1_q$ direction
\begin{align}\label{lastustep}
   -\log\mathcal{Z}_{2,\rm{Weyl}}[{AdS_3\times S^1_q}]&=\frac{1}{4}\int_{-\infty}^{\infty}du\int_0^{\infty}\frac{d\tau}{\sqrt{4\pi\tau^3}} \sum_{n=0}^{\infty}e^{-\frac{\epsilon^2+u^2}{4\tau}}e^{-\tau \frac{n^2}{q^2}}\nonumber\\
   & \times\bigg[\int_{-\infty}^{\infty}d\lambda  \mu^{(3)}_2 (\lambda)(e^{i(\lambda+i)u}+e^{i(\lambda-i)u})+\int_{-\infty}^{\infty}d\lambda  \mu^{(3)}_1 (\lambda)e^{i\lambda u}\bigg]\nonumber\\
   &=\frac{1}{4}\int_{C_{\rm{Odd}}}du\int_0^{\infty}\frac{d\tau}{\sqrt{4\pi\tau^3}} \sum_{n=0}^{\infty}e^{-\frac{\epsilon^2+u^2}{4\tau}}e^{-\tau\frac{n^2}{q^2}}\bigg[W_2^{(3)}( u)(e^u+e^{-u})+W_1^{(3)}(u)\bigg]
\end{align}
The contour $C_{\rm{Odd}}$ is given in the figure \eqref{fig5}.
Substituting \eqref{ftran} in \eqref{lastustep} and performing the sum over Kaluza-Klein modes we obtain
\begin{align}\label{wghyp}
      -\log\mathcal{Z}_{2,\rm{Weyl}}[{AdS_3\times S^1_q}]&=\frac{\log R}{2\pi i}\int_0^{\infty}\frac{du}{2u}\frac{1+e^{-\frac{u}{q}}}{1-e^{-\frac{u}{q}}}  \bigg[\frac{e^{-u} \left(e^u+1\right)^2 \left(-9 e^u+4 e^{2 u}-9 e^{3 u}+4 e^{4 u}+4\right)}{\left(e^u-1\right)^4}\bigg]\nonumber\\
      &=\frac{\log R}{2\pi i}\int_0^{\infty}\frac{du}{2u}\frac{1+e^{-\frac{u}{q}}}{1-e^{-\frac{u}{q}}} \left(+\frac{5 \left(x^3+1\right)}{(1-x)^3}+\frac{5 \left(x^2+x\right)}{(1-x)^3}-\frac{3 \left(x^4+\frac{1}{x}\right)}{(1-x)^3}-\frac{x^4+\frac{1}{x}}{(1-x)^3}\right)
\end{align}
\begin{figure}[h]
\centering
\begin{tikzpicture}[thick,scale=0.85]
\filldraw[magenta] 
                (0,1) circle[radius=3pt]
                (0,-1) circle[radius=3pt];
\draw [decorate,decoration=snake] (0,-1) -- (0,-4);
\draw [decorate,decoration=snake] (0,1) -- (0,4);
\draw[gray,thick] (-2,0.8) node{$\mathbf{\rm{C_{even}}}$};
\draw (0.4,1) node{$\mathbf{\epsilon}$};
\draw (0.4,-1) node{$\mathbf{-\epsilon}$};
\draw[gray, thick] (0,0) -- (0,4);
\draw[gray, thick] (0,0) -- (0,-4);
\draw[gray, thick] (0,0) -- (4,0);
\draw[gray, thick] (0,0) -- (-4,0);
\draw
[
postaction={decorate,decoration={markings , 
mark=at position 0.20 with {\arrow[orange,line width=0.5mm]{>};}}}
][orange, thick] (-4,0.5)--(0,0.5);
\draw[orange, thick] (0,0.5)--(4,0.5);
\draw[gray, thick] (2,3) -- (2,3.4);
\draw[gray, thick] (2,3) -- (2.5,3);
\draw (2.3,3.3) node{$\mathbf{u}$};
\end{tikzpicture}
\caption{Contour $\rm{C_{even}}$ in the $u$-plane for even $AdS_{d}$} \label{fig4}
\end{figure}

 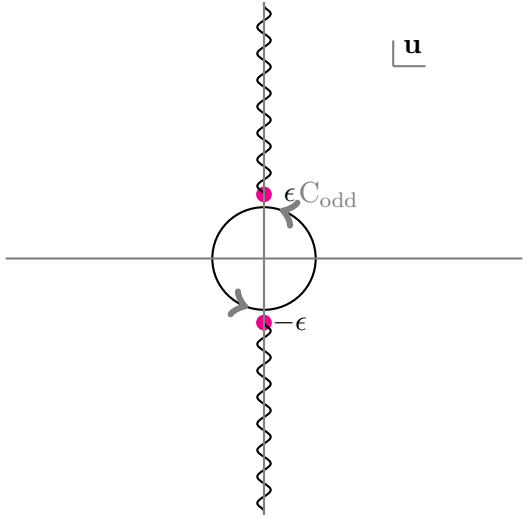
\begin{figure}[h]
\centering
\begin{tikzpicture}[thick,scale=0.85]
\filldraw[magenta] 
                (0,1) circle[radius=3pt]
                (0,-1) circle[radius=3pt];
\draw [decorate,decoration=snake] (0,-1) -- (0,-4);
\draw [decorate,decoration=snake] (0,1) -- (0,4);
\draw
[
postaction={decorate,decoration={markings , 
mark=at position 0.20 with {\arrow[gray,line width=1mm]{>};}}}
]
[
postaction={decorate,decoration={markings , 
mark=at position 0.70 with {\arrow[gray,line width=1mm]{>};}}}
]
(0,0) circle[radius=0.8cm];
\draw [gray,thick](1,1) node{$\mathbf{\rm{C_{odd}}}$};
\draw (0.4,1) node{$\mathbf{\epsilon}$};
\draw (0.4,-1) node{$\mathbf{-\epsilon}$};
\draw[gray, thick] (0,0) -- (0,4);
\draw[gray, thick] (0,0) -- (0,-4);
\draw[gray, thick] (0,0) -- (4,0);
\draw[gray, thick] (0,0) -- (-4,0);
\draw[gray, thick] (2,3) -- (2,3.4);
\draw[gray, thick] (2,3) -- (2.5,3);
\draw (2.3,3.3) node{$\mathbf{u}$};
\end{tikzpicture}
\caption{ Contour $\rm{C_{odd}}$ in the $u$-plane for even $AdS_{d}$} \label{fig5}
\end{figure}
Here $R$ is a dimensionless quantity which is expressed as a ratio of the radial cutoff on $AdS$ to  the radius of $AdS$.\\
Note that at $q=1$, \eqref{wghyp} becomes  the bulk partition function of Weyl-graviton on $S^4$  and it misses out the edge mode partition function. The same property was observed for partition function of conformal $p$-forms \cite{David:2021wrw} and conformal higher derivative fields \cite{Mukherjee:2021alj} on hyperbolic cylinder. The log divergent piece of the partition function can be evaluated by taking residue of \eqref{wghyp} at $u=0$. Therefore the log divergent part of the free energy of Weyl graviton on $S^1_q\times AdS_3$ is given by
\begin{align}
  \mathcal{F}_{2,\rm{Weyl}}^{2,\rm{univ}}(q)=  -\log\mathcal{Z}_{2,\rm{Weyl}}^{2,\rm{univ}}=\frac{553 q^4+70 q^2+1}{60 q^3}\log R
\end{align}
The log divergent piece of the free energy is useful to evaluate the universal contribution of the entanglement entropy of conformal field across a spherical entangling surface. We evaluate the universal contribution of  R\'{e}nyi entropy from the logarithmic divergence of the free energy in an even dimension.
Therefore the  R\'{e}nyi entropy and the entanglement entropy of Weyl graviton in $d=4$ dimension is given by
\begin{eqnarray}
 S_{2,\rm{Weyl}}(q)&=-\frac{(q+1) \left(71 q^2+1\right)}{60 q^3}\log R,\qquad S^{\rm{EE}}_{2,\rm{Weyl}}=-\frac{12}{5}\log R
\end{eqnarray}
In \cite{Huang:2014pfa,David:2020mls}, it was shown that hyperbolic cylinder captures the extractable part of the entanglement entropy for $U(1)$ theory in $d=4$ dimension. It will be nice to have a similar computation demonstrated in \cite{Soni:2016ogt} to show that the extractable part of the entanglement entropy comes from the partition function on hyperbolic cylinder for Weyl graviton.
\subsection{Weyl graviton on $AdS_4$} The structure of the partition function of Weyl graviton on $AdS_4$ is the same as \eqref{partwes4} but the curvature coupled mass terms alter the sign. This is due to the fact that $AdS_4$ has negative curvature. We have given the derivation of the partition function of Weyl graviton on $S^4$ in the appendix \eqref{gaugefixs}. The same analysis will go through for $AdS_4$ altering the sign of the mass terms.  Therefore the partition function can be written as
\begin{align}
    \mathcal{Z}_{2,\rm{Weyl}}[\rm{AdS_4}]&=
  \left(\frac{\det\Delta^{\perp,\rm{AdS_4}}_{1}(3)\det\Delta^{\rm{AdS_4}}_{0}(4)}{\det\Delta^{\perp,\rm{AdS_4}}_{2}(-4)\det\Delta^{\perp,\rm{AdS_4}}_{2}(-2)}\right)^{\frac{1}{2}}
\end{align}
Here $\Delta_{(s)\perp}$ refers to the Laplacian of transverse traceless 
spin-$s$ field on $AdS_{4}$.

To evaluate the partition function on $AdS_4$, we need eigenvalue and degeneracy of spin-$s$ transverse Laplacian on $AdS_4$. The eigenvalues of the transverse traceless Laplacian on $AdS_4$ are given by \cite{Camporesi:1994ga}
\begin{align}\label{adseigen}
    \Delta_{s}^{\perp,\rm{AdS}_4}\psi^{\lambda,u}=(\lambda^2+s+\frac{9}{4})\psi^{\lambda,u}
\end{align}
The Plancherel measure for transverse spin-$s$ field on $AdS_4$  is also known\cite{Camporesi:1994ga}
\begin{align}\label{plmeasure}
    \mu_{(s)}(\lambda)&=\frac{\pi(\lambda^2+(s+\frac{1}{2})^2)}{(2^2\Gamma[2])^2}\lambda \tanh{\pi\lambda}
\end{align}

Using the spectrum and the measure we evaluate the determinant of the massive transverse spin-$s$ field with mass $M_s^2$
\begin{eqnarray}\label{adsstart}
-\frac{1}{2}\log(  {\rm det} \Delta^{\perp,\rm{AdS}_4}_{s} (M_s^2)) = - \frac{1}{2} \int_0^\infty
 d\lambda \mu_s(\lambda) \log \left[  \lambda^2  +(  s+\frac{9}{4})+M_s^2  \right].
\end{eqnarray}
We again replace the logarithm by the identity in (\ref{logiden}) to obtain
\begin{equation}
-\frac{1}{2}\log(  {\rm det}  \Delta^{\perp,\rm{AdS}_4}_{s} (M_s^2)  ) = 
\int_0^\infty \frac{d\tau}{ 2\tau}  \int_0^\infty d\lambda  \mu_s (\lambda) ( 
e^{-\tau ( \lambda^2 + ( \nu_s)^2 ) } - e^{-\tau} ) .
\end{equation}
where $\nu_s^2=M_s^2+s+\frac{9}{4}$. Now we need to evaluate the integral over Plancherel measure in order to evaluate the last term of the previous line. 
 We use the fact that the Plancherel measure is symmetric in $\lambda$. 
The integral over $\lambda$ in Plancherel measure can be written in terms of its Fourier transform. 
\begin{eqnarray}
\mu_s^{(d)}(\lambda) = \int_C \frac{du}{2\pi} e^{ - i \lambda u } W_s^{(d)}( u ) .
\end{eqnarray}
 Fourier transform is well defined in the contour $C$.  The contour differs in even and odd dimensions. From \eqref{ftran}, one can observe that $\lim_{u\rightarrow 0}W_s^{(d)}(u)=0$ which implies that the integral over Planchrerel measure of transverse spin-$s$ field vanishes, which is 
 \begin{align}
     \int_0^{\infty}d\lambda \mu_s^{(d)}(\lambda)=0
 \end{align}
 Therefore we proceed with the first term and perform the integration over $\lambda$, 
we obtain 
\begin{equation}
-\frac{1}{2}\log(  {\rm det} \Delta^{\perp,\rm{AdS}_4}_{s}  (M_s^2)  )   = \int_C  du
\int_0^\infty \frac{d\tau}{ 8 ( \pi \tau^3)^{\frac{1}{2}}  }
e^{ - \frac{\epsilon^2 + u^2}{4\tau} - \tau   ( \nu_s)^2  } W_s^{(4)}( u ) .
\end{equation}
The expression of $ W_s^{(4)}( u )$ can be obtained from \eqref{ftran}.
After integration over $\tau$ we obtain 
\begin{equation} \label{adsend}
-\frac{1}{2}\log(  {\rm det}  \Delta^{\perp,\rm{AdS}_4}_{s}  (M_s^2)  )     = 
\int_C \frac{du}{ 4 \sqrt{ u^2 + \epsilon^2}} e^{  - \nu_s \sqrt{ u^2 + \epsilon^2}  }  W_s^{(4)}(u) 
\end{equation}
We can take the $\epsilon\rightarrow 0$ limit at the end and substitute \eqref{ftran} to obtain the logarithm of the determinants.

We now obtain
\begin{align}\label{adsall}
\begin{split}
    -\frac{1}{2}\log(  {\rm det} \Delta^{\perp,\rm{AdS}_4}_{2}  (-4)  )   &=\int_C\frac{du}{2u}\frac{1+x}{1-x}\left(5 \frac{x^2}{(1-x)^3}-5\frac{x}{1-x}\right)\\
   -\frac{1}{2}\log(  {\rm det}  \Delta^{\perp,\rm{AdS}_4}_{2}(-2)   )   &=\int_C\frac{du}{2u}\frac{1+x}{1-x}\left(5\frac{x^3}{(1-x)^3}-5\frac{x^2}{1-x}\right)\\
    -\frac{1}{2}\log(  {\rm det}  \Delta^{\perp,\rm{AdS}_4}_{1}  (3)  )   &=\int_C\frac{du}{2u}\frac{1+x}{1-x}\left(\frac{3 x^4}{(1-x)^3}-\frac{x^3}{1-x}\right)\\
     -\frac{1}{2}\log(  {\rm det} \Delta^{\rm{AdS}_4}_{0}  (4)  )   &=\int_C\frac{du}{2u}\frac{1+x}{1-x}\frac{x^4}{(1-x)^3}
\end{split}
\end{align}
Here $x=e^{-u}$. To obtain the complete bulk and edge character, we assemble all the terms given in \eqref{adsall} and write the partition function as integral over characters.  Therefore the total `naive' bulk and the edge character of Weyl graviton on $AdS_4$ is given by
\begin{align}
    \chi^{b,\rm{AdS}}_{2,\rm{Weyl}}&=\left(\frac{5 \left(x^2\right)}{(1-x)^3}+\frac{5 \left(x^3\right)}{(1-x)^3}-\frac{3 \left(x^4\right)}{(1-x)^3}-\frac{x^4}{(1-x)^3} \right)\\
     \chi^{e,\rm{AdS}}_{2,\rm{Weyl}}&=\left(\frac{5 (x)}{1-x}+\frac{5 \left(x^2\right)}{1-x}-\frac{x^3}{1-x}\right)
 \end{align}
 Note that there are no negative powers of $x$ in the numerator and therefore it is a unitary irreducible representation of $SO(2,3)$. These expressions of each transverse character agree with the character of a massive transverse spin-$2$ field given in\cite{Sun:2020ame}. Here the mass terms are specific which arise from the curvature coupling.

 \section{Free Conformal higher spin field in $d=4$ dimension}

 In this section, we generalize the method to express the partition function as integral over characters for conformal higher spin fields in the $d=4$ dimension. We express the partition function of conformal higher spin fields on $S^4$ and on $AdS_4$ in terms of `naive' bulk and edge characters. We evaluate the partition function on the hyperbolic cylinder as well and observed that it captures only the bulk mode of the partition function and misses out  the edge mode. The similar property was also observed for conformal $p$-forms  \cite{David:2021wrw} and for conformal higher derivative fields  \cite{Mukherjee:2021alj}.
 \subsection{Free Conformal higher spin field on $S^4$}
 We study the partition function of conformal higher spin fields on $S^4$. We express the partition function as integral over characters. The strategy is to first express the logarithm of the determinant of  the massive transverse field in terms of integral over characters and finally combine all the ghosts to obtain the partition function.
 
 The partition function of free conformal higher spin fields on $S^4$ is given by \cite{Giombi:2014yra}
 \begin{align}
     \mathcal{Z}^{(S^4)}_{s,\rm{Weyl}}&=\prod_{k=0}^{s-1}\left(\frac{\det\left(-\nabla^2+k-(s-1)(s+2)\right)_{k\perp}}{\det\left(-\nabla^2+s-(k-1)(k+2)\right)_{s\perp}}\right)^{\frac{1}{2}}.
 \end{align}
 We now follow the same prescription which we developed for Weyl graviton on $S^4$ to express the partition function in terms of `naive' bulk and edge characters.  We write the logarithm of the partition function
 \begin{align}
     \log\mathcal{Z}^{(S^4)}_{s,\rm{Weyl}}&=\frac{1}{2}\sum_{k=0}^{s-1}\left(\log\det\Delta^{\perp}_k(M^2_k)-\log\det\Delta^{\perp}_s(M^2_s)\right).
 \end{align}
 Here $\Delta^{\perp}_s(M^2_k)$ is the transverse spin-$s$ Laplacian on $S^4$ with mass term $M_s^2$. Note that the numerator of the partition function comes from the `ghost' contribution and one gets all ghost contributions starting from spin-$0$ to spin-$(s-1)$.
 We use the eigenvalue and degeneracy of the transverse symmetric spin-$s$ tensor field given in \eqref{egdeg} and follow \eqref{repstep} to \eqref{parttensor} to obtain
 \begin{align}
 \begin{split}
      \frac{1}{2}\sum_{k=0}^{s-1}\log\det\Delta^{\perp}_k(M^2_k)  &=-\sum_{k=0}^{s-1}\int_{\epsilon}^{\infty}\frac{dt}{2\sqrt{t^2-\epsilon^2}}\bigg[\left(e^{i\nu_k\sqrt{t^2-\epsilon^2}}+e^{-i\nu_k\sqrt{t^2-\epsilon^2}}\right)f^{(k)} (i t)\bigg],\\
       \frac{1}{2}\sum_{k=0}^{s-1}\log\det\Delta^{\perp}_k(M^2_s)&=\sum_{k=0}^{s-1}\int_{\epsilon}^{\infty}\frac{dt}{2\sqrt{t^2-\epsilon^2}}\bigg[\left(e^{i\nu_s\sqrt{t^2-\epsilon^2}}+e^{-i\nu_s\sqrt{t^2-\epsilon^2}}\right)f^{(s)} (i t)\bigg],
 \end{split}
 \end{align}
 where $f^{(s)}(it)$ and $\nu_s$ are given by 
 \begin{align}
     f^{(s)}(it)=\sum_{n=-1}^{\infty}g^{(s)}_n e^{-t(n+\frac{3}{2}}),\qquad \nu_s=i(s+\frac{1}{2}).
 \end{align}
We now perform the sum and
 combine the numerator and the denominator to obtain
 \begin{align}\label{sps}
-\log\mathcal{Z}^{(S^4)}_{s,\rm{Weyl}}&=\sum_{k=0}^{s-1}\int_0^{\infty}\frac{1+x}{1-x}\bigg[\frac{(2 s+1) \left(x^{\frac{3}{2}-\left(k+\frac{1}{2}\right)}+x^{\left(k+\frac{1}{2}\right)+\frac{3}{2}}\right)}{(1-x)^3}-\frac{(2 k+1) \left(x^{\frac{3}{2}-\left(s+\frac{1}{2}\right)}+x^{\left(s+\frac{1}{2}\right)+\frac{3}{2}}\right)}{(1-x)^3}\nonumber\\
&-\frac{s (s+1) (2 s+1) \left(x^{\frac{1}{2}-\left(k+\frac{1}{2}\right)}+x^{\left(k+\frac{1}{2}\right)+\frac{1}{2}}\right)}{6 (1-x)}+\frac{k (k+1) (2 k+1) \left(x^{\frac{1}{2}-\left(s+\frac{1}{2}\right)}+x^{\left(s+\frac{1}{2}\right)+\frac{1}{2}}\right)}{6 (1-x)}\bigg]\nonumber\\
&=\sum _{k=0}^{s-1}\int_0^{\infty}\frac{dt}{2t} \frac{1+x}{1-x} \bigg[ \chi^b_{s,\perp}- \chi^b_{k,\perp}-\chi^e_{s,\perp}+\chi^e_{k,\perp}\bigg].
 \end{align}
 Here $x=e^{-t}$ and $\chi^b_{s,\perp}$, $\chi^e_{s,\perp}$ represents the bulk and the edge characters of massive transverse symmetric field respectively. The partition function of free conformal higher spin field can now be expressed as integral over
 the `naive' bulk and edge characters. 
 \begin{align}\label{chsch}
    \chi^{b,\rm{dS}}_{s,\rm{Weyl}}&=\sum_{k=0}^{s-1}\bigg[-\frac{(2 k+1) \left(x^{\frac{3}{2}-\left(s+\frac{1}{2}\right)}+x^{\left(s+\frac{1}{2}\right)+\frac{3}{2}}\right)}{(1-x)^3}+\frac{(2 s+1) \left(x^{\frac{3}{2}-\left(k+\frac{1}{2}\right)}+x^{\left(k+\frac{1}{2}\right)+\frac{3}{2}}\right)}{(1-x)^3}\bigg]\\
      \chi^{e,\rm{dS}}_{s,\rm{Weyl}}&=\sum_{k=0}^{s-1}\bigg[\frac{s (s+1) (2 s+1) \left(x^{\frac{1}{2}-\left(k+\frac{1}{2}\right)}+x^{\left(k+\frac{1}{2}\right)+\frac{1}{2}}\right)}{6 (1-x)}\nonumber\\
      &\qquad\qquad\qquad\qquad\qquad\qquad-\frac{k (k+1) (2 k+1) \left(x^{\frac{1}{2}-\left(s+\frac{1}{2}\right)}+x^{\left(s+\frac{1}{2}\right)+\frac{1}{2}}\right)}{6 (1-x)}\bigg]
 \end{align}
 One obtains the log coefficient of the free energy by expanding the integrand \eqref{sps} about $t=0$ and collecting the $1/t$ coefficient. Therefore we obtain the log coefficient of free conformal higher spin fields in $d=4$ dimension, which agrees with \cite{Tseytlin:2013fca,Anninos:2020hfj}
 \begin{align}\label{unt}
     \alpha^{(4)}_s=-\frac{1}{180} s^2 (s+1)^2 \left(14 s^2+14 s+3\right)
 \end{align}
We finally obtain the partition function of conformal higher spin fields on $S^4$ as an integral over `naive' characters and extract the universal or logarithmic divergent part. But note that, these `naive' characters grow as $e^{t}$ and therefore can not be a unitary irreducible representation of $SO(1,4)$. So one has to replace it with the flipped character \cite{Anninos:2020hfj}. But for our purposes, the `naive' characters are enough to compare the partition functions.
\subsection{Free conformal higher spins on $AdS_3\times S^1_q$} \label{spinee}
In \cite{Beccaria:2014jxa} it was shown that the $2s$-derivative conformal spin-$s$ operator, evaluated
on the $S^1\times S^3$ background and restricted to transverse traceless spin $s > 0$ tensors  takes the following form:
\begin{eqnarray}\label{volads}
\mathcal{O}^{S^1\times S^3}_{2s} = 
\begin{cases}
\prod_{p=1}^{s}\left((\partial_0+2p-s-1)^2-\Delta_s-s-1\right)\qquad  & \hbox{for}\;  s \; \hbox{even} ,  \\
\prod_{p=-\frac{s-1}{2}}^{\frac{s-1}{2}}\left((\partial_0+2p)^2-\Delta_s-s-1\right)
\qquad  & \hbox{for} \; s\; \hbox{odd}.
\end{cases}
\end{eqnarray}
It follows the same pattern of the kinetic operator of conformal spin-$1$ and Weyl graviton on $S^1\times S^3$ \cite{Beccaria:2014jxa} and therefore it was generalized for conformal higher spin fields. 
And the proposal for the partition was
\begin{align}\label{partassum}
    \mathcal{Z}_{s,\rm{Weyl}}[S^1\times S^3]&=\frac{1}{\left(\prod_{k=1}^s\mathcal{O}_{k,\perp}\right)^{\frac{1}{2}}}
\end{align}

But as far as we are aware the general form of the action of conformal higher spin field is not known yet.
 So we follow the same prescription to obtain the partition function on the background $S^1\times AdS_3$ because the procedure to obtain the partition function of spin-$1$ and Weyl graviton is the same on $S^1\times S^3$ and $S^1\times AdS_3$. The only difference in the partition function is in the sign of the mass term because of the negative curvature of the $AdS$ space. The form of the partition function remains the same but the curvature coupled mass alters the sign. Note that, the form of the partition function (barring the sign of the mass term) of conformal vector field on $S^1\times S^3$ \cite{Beccaria:2014jxa} is same as the partition function on $S^1\times AdS_3$ \cite{David:2020mls} and it is also same for Weyl graviton on $S^1\times AdS_3$ which is given in \eqref{weylpart} and on $S^1\times S^3$ given in\cite{Beccaria:2014jxa}. Therefore the kinetic operator of the conformal higher spin field on $S^1\times AdS_3$  will also have the similar form given in \eqref{volads} but the mass term will alter its sign. 
\begin{eqnarray}\label{volads}
\mathcal{O}^{S^1\times AdS_3}_{2s} = 
\begin{cases}
\prod_{p=1}^{s}\left(\partial_0^2-\left((-\Delta_s-s-1)^{\frac{1}{2}}+i(2p-s-1)\right)^2\right)\qquad  & \hbox{for}\;  s \; \hbox{even} ,  \\
\prod_{p=-\frac{s-1}{2}}^{\frac{s-1}{2}}\left(\partial_0^2-\left((-\Delta_s-s-1)^{\frac{1}{2}}+i 2p)\right)^2\right)
\qquad  & \hbox{for} \; s\; \hbox{odd}.
\end{cases}
\end{eqnarray}
Note that this formula is consistent with the kinetic operator of Maxwell field \cite{David:2020mls} and Weyl graviton  on $S^1\times AdS_3$ \eqref{weylpart}.
Therefore the form of the partition function also takes the similar form given in \eqref{partassum} and we follow the same steps to express the partition function in terms of the integral over characters. 
\begin{align}
    -\log\mathcal{Z}_{s,\rm{Weyl}}[S^1_q\times AdS_3]&=\frac{1}{2}\sum_{k=1}^{s}\log\mathcal{O}_{k,\perp}
\end{align}

Since the operator \label{volads} takes different forms for even and odd spin cases, we separate the odd and even spin operators and finally add them to evaluate the partition function.
For even spin,we follow the same steps from \eqref{repstep} to \eqref{parttensor} to obtain
\begin{align}
    \log\mathcal{O}_{2k,\perp}&=\frac{1}{2}\sum_{p=1}^{2k}\int_C\frac{du}{2u}W_{2k}^{(3)}(u)e^{-u(2p-2k-1)}\left(1+2\sum_{n=1}^{\infty}e^{-u\frac{n}{q}}\right)
\end{align}
For odd spin case, we obtain
\begin{align}
      \log\mathcal{O}_{2k-1,\perp}&=\frac{1}{2}\sum_{p=-(k-1)}^{k-1}\int_C\frac{du}{2u}W_{2k-1}^{(3)}(u)e^{-u(2p)}\left(1+2\sum_{n=1}^{\infty}e^{-u\frac{n}{q}}\right)
\end{align}
Here $W_{2k}^{(3)}(u)$ is the Fourier transform of the Plancherel measure of transverse spin-$2k$ field on $AdS_3$ which can be obtained from the equation \eqref{ftran}.
Now we add up the even and odd spin contributions to obtain the partition function
\begin{align}\label{hyps}
    -\log\mathcal{Z}_{s,\rm{Weyl}}[AdS_3\times S^1_q]&=\sum_{k=1}^{\frac{s}{2}}  \left(  \log\mathcal{O}_{2k}+ \log\mathcal{O}_{2k-1}\right)\nonumber\\
    &=\frac{\log R}{2\pi i}\int_C\frac{du}{2u}\frac{1+e^{-\frac{u}{q}}}{1+e^{-\frac{u}{q}}}\nonumber\\
    &\times\sum _{k=0}^{s-1} \left(\frac{(2 s+1) \left(x^{\frac{3}{2}-\left(k+\frac{1}{2}\right)}+x^{\left(k+\frac{1}{2}\right)+\frac{3}{2}}\right)}{(1-x)^3}-\frac{(2 k+1) \left(x^{\frac{3}{2}-\left(s+\frac{1}{2}\right)}+x^{\left(s+\frac{1}{2}\right)+\frac{3}{2}}\right)}{(1-x)^3}\right)
\end{align}
Note that the sum over Kaluza-Klein modes give the kinetic factor $\frac{1+e^{-\frac{u}{q}}}{1+e^{-\frac{u}{q}}}$ in the partition function.
The character of conformal higher spin fields on $S^1_q\times AdS_3$ includes all the characters of massive transverse spins starting from spin-$0$ to $s-1$ where masses are determined from the curvature of $AdS_3$.

We obtain the partition function or the free energy as integral over characters but note that this is the same as the `naive' bulk character of the conformal spin-$s$ field on $S^4$ given in \eqref{chsch}.  Therefore the partition function of conformal higher spin fields on $S^1\times AdS_3$ captures only the bulk character and misses out the edge part. 
\paragraph{Free energy and entanglement entropy}

The contour $C$ for odd $AdS$ is given in figure \eqref{fig5}. So to obtain the universal contribution of the partition function, we evaluate the residue of the integral \eqref{hyps} at $u=0$.
\begin{align}
    \mathcal{F}_{q,s}[AdS_3\times S^1_q]^{\rm{univ}}&=-\frac{s (s+1) \left(q^4 (2 s (s+1) (9 s (s+1)-7)-11)+10 q^2 \left(s^2+s+1\right)+1\right)}{360 q^3}
\end{align}
We also evaluate the universal contribution of the   entanglement entropy of conformal higher spin fields across a spherical entangling surface
\begin{align}\label{eechs}
    S_{s,\rm{Weyl}}(q)&=-\frac{(q+1) s (s+1) \left(q^2 (10 s (s+1)+11)+1\right)}{360 q^3}\log R,\\ S^{\rm{EE}}_{s,\rm{Weyl}}&=-\frac{1}{90} s (s+1) (5 s (s+1)+6)\log R
\end{align}
With the substitution of $s=1$, one obtains the entanglement entropy of conformal spin-$1$ field in $d=4$ dimension
which agrees with the extractable entanglement entropy of Maxwell field in $d=4$ dimension \cite{Donnelly:2014fua,Soni:2016ogt,Casini:2015dsg}. In \cite{David:2021wrw}, it was shown that the extractable part of the entanglement entropy of Maxwell field in $d=4$ dimension comes from the partition function of the hyperbolic cylinder.  It will be nice to find the relation between the non-extractable part of entanglement entropy and the edge mode partition function of Weyl graviton and conformal higher spin fields in $d=4$ dimension.
\subsection{Free conformal higher spins on $AdS_4$}
In this section, we express the partition function of conformal higher spin fields in terms of the integral over characters on $AdS_4$. We first express the logarithm of the determinant of the transverse traceless field on $AdS_4$ and finally combine all the ghost contributions to write the full partition function.
 The partition function of free conformal higher spin fields on $AdS_4$ is given by \cite{Klebanov:2011uf}.
 \begin{align}\label{partads4}
     \mathcal{Z}_{s,\rm{Weyl}}[AdS_4]&=\prod_{k=0}^{s-1}\left(\frac{\det\left(-\nabla^2-k+(s-1)(s+2)\right)_{k\perp}}{\det\left(-\nabla^2-s+(k-1)(k+2)\right)_{s\perp}}\right)^{\frac{1}{2}}.
 \end{align}
 Note that this partition function has a similar expression as the partition function of conformal higher spin fields on $S^4$, but the mass terms have altered the sign. This is because of the fact that $AdS$ is the Einstein space with negative curvature and for the conformal fields, mass terms come from the curvature coupling.
  We now follow the similar procedure described for the Weyl graviton on $AdS_4$ to express the partition function in terms of `naive' bulk and edge characters. Therefore we write the logarithm of the partition function
 \begin{align}
     \log\mathcal{Z}_{s,\rm{Weyl}}[AdS_4]&=\frac{1}{2}\sum_{k=0}^{s-1}\left(\log\det\Delta^{\perp}_k(M^2_k)-\log\det\Delta^{\perp}_s(M^2_s)\right).
 \end{align}
 Here $\Delta^{\perp}_s(M^2_s)$ is the transverse spin-$s$ Laplacian on $AdS_4$ with mass $M_s^2$. The mass term can be read off from the denominator of  \eqref{partads4}. Note that the numerator of the partition function comes from the 'ghost' contribution and one gets all ghost contributions starting from spin-$0$ to spin-$(s-1)$.
 
 To evaluate the partition function we need to use the eigen-value and Plancherel measure of the transverse symmetric spin-$s$ tensor field given in \eqref{adseigen},\eqref{plmeasure} and follow \eqref{adsstart} to \eqref{adsend} to obtain
 \begin{align}
 \begin{split}
      \frac{1}{2}\sum_{k=0}^{s-1}\log\det\Delta^{\perp}_k(M^2_k)  &=-\sum_{k=0}^{s-1}\int_{C}\frac{du}{4\sqrt{u^2+\epsilon^2}}e^{-\nu_k u}W_k^{(4)}(u)\\
       \frac{1}{2}\sum_{k=0}^{s-1}\log\det\Delta^{\perp}_s(M^2_s)&=-\sum_{k=0}^{s-1}\int_{C}\frac{du}{4\sqrt{u^2+\epsilon^2}}e^{-\nu_s u}W_s^{(4)}(u)
 \end{split}
 \end{align}
 $W_s^{(4)}$ is the Fourier transform of the Plancherel measure of the transverse spin-$s$ field on $AdS_4$ and the expression can be found in \eqref{ftran}.
We now combine all the ghost contributions and obtain the partition function
 \begin{align}
     -\log\mathcal{Z}_{s,\rm{Weyl}}[AdS_4]&=\sum_{k=0}^{s-1}\int_{C}\frac{du}{4\sqrt{u^2+\epsilon^2}}\left(e^{-\nu_s u}W_s^{(4)}(u)-e^{-\nu_k u}W_k^{(4)}(u)\right)\nonumber\\
     &=\sum_{k=0}^{s-1}\int_{C}\frac{du}{4\sqrt{u^2+\epsilon^2}}\frac{1+x}{1-x}\bigg[\left((2s+1)\frac{x^{k+2}}{(1-x)^3}-\frac{1}{6} s (s+1) (2 s+1)\frac{x^{k+1}}{1-x}\right)\nonumber\\
     &-\left((2k+1)\frac{x^{s+2}}{(1-x)^3}-\frac{1}{6} k (k+1) (2 k+1)\frac{x^{s+1}}{1-x}\right)\bigg]
 \end{align}
 Here $x=e^{-u}$ and the contour $C$ is given in the picture \eqref{fig4} which runs over the entire real line. Therefore we express the partition function in terms of the integral over characters. One can read off the bulk and the edge characters from the partition function 
 \begin{align}
    \chi^{b,\rm{AdS}}_{s,\rm{Weyl}}&=\sum_{k=0}^{s-1}\bigg[\left((2s+1)\frac{x^{k+2}}{(1-x)^3}\right)-\left((2k+1)\frac{x^{s+2}}{(1-x)^3}\right)\bigg]\\
     \chi^{b,\rm{AdS}}_{s,\rm{Weyl}}&=\sum_{k=0}^{s-1}\bigg[\frac{1}{6} s (s+1) (2 s+1)\frac{x^{k+1}}{1-x}-\frac{1}{6} k (k+1) (2 k+1)\frac{x^{s+1}}{1-x}\bigg]
 \end{align}
  Note that there are no negative powers of $x$ in the numerator and therefore we not need to `flip' the character. These expressions of each transverse character agree with the character of a massive transverse spin-$s$ field given in \cite{Sun:2020ame}. Here the mass terms are specific which arise from the curvature coupling.
  
To obtain the universal term of the partition function, one expands the integrand and collects the $1/u$ term which contributes to the logarithmic divergent or universal part of the partition function.
 \begin{align}
 -\log \mathcal{Z}_{s,\rm{Weyl}}[AdS_4]^{\rm{univ}}&=\sum_{k=0}^{s-1}\frac{1}{360} (k-s) \big(30 k^3 s+15 k^3+90 k^2 s^2+135 k^2 s+50 k^2\nonumber\\
 &+30 k s^3+135 k s^2+140 k s+45 k+15 s^3+50 s^2+45 s+14\big)\nonumber\\
 &=-\frac{1}{360} s^2 (s+1)^2 (14 s (s+1)+3)
 \end{align}
Note that the logarithmic divergent piece or universal term in the partition function  of conformal higher spin fields on $AdS_4$ is the half of that obtained from the partition function on $S^4$ given in \eqref{unt}. Similar observations were also made for co-exact $p$-forms and therefore the partition function of $p$-forms in even $AdS$ becomes  half of the partition function on even sphere \cite{David:2020mls}.

One can understand this property from the trace  anomaly on the even dimensional $AdS_d$. 
Free energies become proportional to the
trace anomaly coefficient $a_{d}$ in conformally flat background. The expectation value of the trace of the stress tensor is given by \cite{Herzog:2017xha}
\begin{eqnarray}
\langle T^\mu_\mu \rangle = \frac{1}{ (4\pi )^{\frac{d}{2}}} 
\left( \sum_j c_{(d) j}  I_{j+1}^{(d)}  - ( -1)^{ \frac{ ( d)}{2} } a_{d} E_{d} \right) .
\end{eqnarray}
Here $E_{d}$ is the Euler density and the expression is given by 
\begin{eqnarray} \label{euden}
E_{d}  = \frac{1}{ 2^{ \frac{d}{2}}} \delta^{\nu_1\cdots \nu_{d}}_{\mu_1\cdots \mu_{d}} 
R^{\mu_1\mu_2}_{\;\;\;\nu_1\nu_2} \cdots R^{\mu_d \mu_{d}}_{\;\;\;\nu_d \nu_{d}} 
\end{eqnarray}
 and $I_j^{d}$ are independent Weyl invariants of weight 
 $- d$. The variation of the partition function with respect to metric which is conformally equivalent to a flat metric in $d=4$ dimension is known \cite{Brown:1977sj,Herzog:2013ed}
\begin{eqnarray}
\frac{2}{ \sqrt{g} } \frac{\delta \log {\cal Z}_{s,\rm{Weyl}}}{ \delta g_{\mu\nu} } &=& 
\langle T^{\mu\nu} \rangle , \\ \nonumber
&=& -\frac{a_4}{(4\pi)^2}\left(g^{\mu\nu}(\frac{R^2}{2}-R_{\lambda\rho}^2)+2R^{\mu\lambda}R^{\nu}_{~\lambda}-\frac{4}{3}RR^{\mu\nu}\right).
\end{eqnarray}
Now one can integrate the above equation to get the free energy
\begin{equation}
\log {\cal Z}_{s\rm{Weyl}} = a_4  \int d^4 x \sqrt{g} \mathcal{K} ( R^{(2)} ) .
\end{equation}
Note that $\mathcal{K}(R^{(2)} )$, is the quadratic invariant function of the curvature tensor.
But the curvature tensor of $AdS$ and sphere differs only by a sign and therefore $\mathcal{K}(R^2)$ remains same in both cases. But the integral over measure gives a factor which is proportional to the volume and therefore the ratio of the partition functions becomes the ratio of the volumes of $AdS_4$ and $S^4$. We use the regularised volume of $AdS_4$ given in \cite{Hung:2014npa} to obtain
\begin{eqnarray}
\frac{ \log {\cal Z}_{s,\rm{Weyl}}[AdS_4]}{ \log{\cal Z}_{s,\rm{Weyl}} [S^4]} = 
\frac{ {\rm Vol } ( AdS_4) } { {\rm Vol } ( S^4)} 
&=& \frac{1}{2} .
\end{eqnarray}
 We observe that this is true even for the conformal scalars. The universal term of the partition function of conformal scalar on $S^4$ is known $-\frac{1}{90}$ \cite{Anninos:2020hfj}. Similarly one can also extract the universal term in the partition function on $AdS_4$ which is given by  $\alpha_{4,0}^{\rm{AdS}}=\frac{\nu^4}{24}-\frac{\nu^2}{48}-\frac{17}{5760}$ \cite{Sun:2020ame}. For conformal scalar on $AdS_4$, $\nu=\frac{1}{2}$ and therefore one obtains $\alpha_{4,0}^{\rm{AdS}}=-\frac{1}{180}$, which is half of the conformal scalar on $S^4$.
 \section{Hofman-Maldacena variables and causality bound of conformal higher spins}\label{causality}
 
 Conformal higher spin fields have higher derivative kinetic terms in the action. The higher derivative kinetic term will have a negative residue in the propagator which indicates a negative norm state \cite{Donoghue:2017fvm}. Therefore one expects the conformal higher spin theories are also non-unitary. In this section, we use averaged null energy condition to show that this class of theories are indeed non-unitary.
 
 From the free energy of conformal higher spin fields on the hyperbolic cylinder, we evaluate the conformal dimension of the co-dimension two twist operator \cite{Hung:2014npa}.
 \begin{align}
     h_{q,s}&=\frac{q}{3\rm{Vol}(AdS_{3})}\left(\partial_q\mathcal{F}_s[AdS_3\times S^1_q]|_{q=1}-\partial_q\mathcal{F}_s[AdS_3\times S^1_q]\right)\nonumber\\
     &=\frac{\left(q^2-1\right) s (s+1) \left(q^2 (10 s (s+1)+13)+3\right)}{2160 \pi  q^3}
 \end{align}
 The regularised volume of $AdS_{3}$  is  given in \eqref{regv} \cite{Hung:2014npa}.
 The conformal dimension of the twist operator can be thought of as the energy density across the entangling surface. The first and second derivatives of $h_{q,s}$ with respect to $q$, at $q=1$, can be expressed as two and three-point functions of the stress tensor \cite{Hung:2014npa}. Here $2\pi q$ is the inverse temperature of the thermal ensemble. In conformal field theory, the two and three-point functions of the stress tensor are characterized by the parameters $a$, $b$ and $c$ \cite{Osborn:1993cr}
 and these parameters  obey the following relations 
\begin{eqnarray}\label{hqabc}
& & h_q^\prime|_{q=1}=\frac{\pi ^{\frac{d+3}{2}} C_T}{2^{d-3} d \left(d^2-1\right) \Gamma \left(\frac{d-1}{2}\right)} , \quad
C_T =\frac{\left(8  \pi ^{d/2}\right) (a (d-2) (d+3)-2 b-c (d+1))}{d (d+2) \Gamma \left(\frac{d}{2}\right)},
\nonumber \\
& &  \nonumber
h_q^{\prime\prime} |_{q=1} =-\frac{16 \pi ^{d+1} }{d^2 \Gamma (d+3)}
 \left[2 a \left(3 d^2-3 d-4\right) (d-2)-2 b d (d-1)-c (3 d-4) (d+1)\right],
 \\
\end{eqnarray}
Here $C_T$ is the central charge  and $a,b,c$  are the parameters determining the $3$-point function of the stress tensors \cite{Osborn:1993cr}. We have two relations and three parameters and therefore we can not determine them independently but we find a linear relation that will help us to determine the unitarity of the theory.

We will use the positivity of the energy flux \cite{Hofman:2008ar} or average null energy condition to show the non-unitarity of the theory.
 Hofman-Maldacena variables $t_2$ and $t_4$  were constructed from the ratio of the combinations of the three parameters $a$, $b$, and $c$ to determine the positive energy of the flux. The relations are known for arbitrary dimensions \cite{Buchel:2009sk}
\begin{figure}
\centering
\begin{subfigure}{.5\textwidth}
  \centering
  \includegraphics[width=1\linewidth]{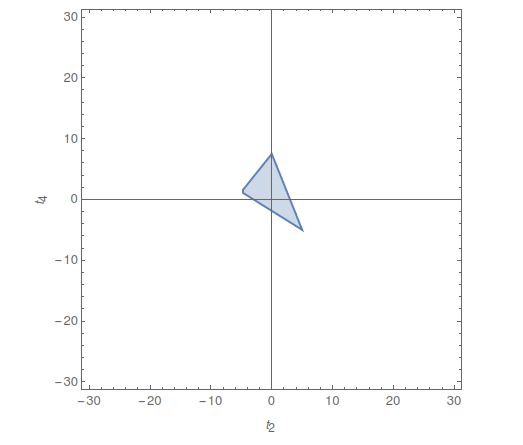}
  \caption{Region of unitarity in $d=4$ dimension.}
  \label{fig:vec1}
\end{subfigure}%
\begin{subfigure}{.5\textwidth}
  \centering
  \includegraphics[width=1\linewidth]{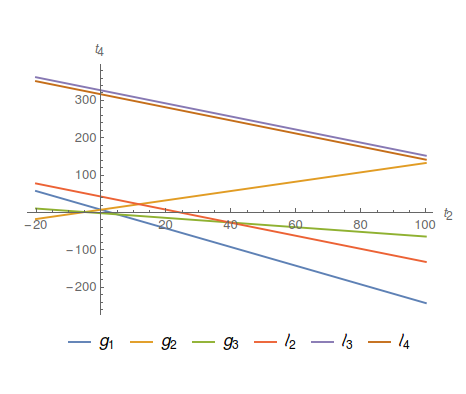}
  \caption{CHS in $d=4$ dimension.}
  \label{fig:fer1}
\end{subfigure}
\label{fig:test}
\caption{The region of unitarity is the small triangle where $g_1$, $g_2$ and $g_3$ meet near the origin. The region near the origin is shown clearly in the left picture.}
\end{figure}
\begin{align}
    \begin{split}
     t_2&=\frac{2 (d+1) (a (d-1) ((d+8) d+4)-d (-3 b d+2 c d+c))}{d (a (d-2) (d+3)-2 b-c (d+1))},\\
     t_4&=\frac{(d+1) (d+2) \left(3 a \left(-2 d^2+d+1\right)+d (-2 b d+c d+c)\right)}{d (a (d-2) (d+3)-2 b-c (d+1))}.
    \end{split}
\end{align}
The positivity of energy flux puts the bound on the parameters $t_2$ and $t_4$ for conformal field theories in $d>3$ dimensions and they are given by \cite{Camanho:2009vw} 
\begin{align}\label{bounds}
    \begin{split}
    g_1&=1 -\frac{t_2}{d-1}-\frac{2t_4}{(d+1) (d-1)}\geq 0,\\
  g_2 &=1 -\frac{t_2}{d-1}-\frac{2 t_4}{(d-1) (d+1)}+\frac{t_2}{2}\geq 0,\\
 g_3&=1+  \frac{(d-2) (t_2+t_4)}{d-1}-\frac{t_2}{d-1}-\frac{2 t_4}{(d-1) (d+1)}\geq 0.
    \end{split}
\end{align}
The bounds on the parameters $t_2$ and $t_4$ represent a triangular region in the $t_2-t_4$-plane and all the unitary conformal field theories are points inside the region. Here our strategy is to find a linear relation of the Hofman-Maldacena variables which will represent a straight line in $t_2-t_4$ plane. If the straight line never intersects the region at any point, we can say that the theory which is  a point on the straight line will never be inside the triangular region. Therefore we can conclude that the theory does not satisfy the unitarity constraint.

We first investigate it for Weyl graviton in $d=4$ dimension. The conformal dimension of the twist operator and its first and second derivative at $q=1$ are given by
\begin{align}
     h_{q,s=2}=\frac{\left(q^2-1\right) \left(73 q^2+3\right)}{360 \pi  q^3},\qquad h_{q,s=2}^{\prime}=\frac{19}{45 \pi },\qquad h_{q,s=2}^{\prime\prime}=-\frac{22}{45 \pi }.
\end{align}
From the first and second derivative of the twist operator at $q=1$,  we find a linear relation of the variables $t_2$ and $t_4$ which is a straight line on the plane. The equation of the straight line is given by
\begin{align}
    \ell _2=t_4-\frac{1}{76} \left(3246-133 t_2\right)=0.
\end{align}
The intercepts are given by $t_2=24.406$ and $t_4=42.7105$ and it never intersects the triangular region formed by $g_1$, $g_2$ and $g_3$. 

Similarly, we analyze conformal higher spin theories ($s\geq 2$) and observe that the straight line in the $t_2-t_4$ plane never intersects the region of unitarity. In the figure \eqref{fig:fer1}, we show it explicitly for conformal spin-$3$ and conformal spin-$4$ cases.

From the above analysis, we verify that conformal higher spin theories do not satisfy the positivity of energy flux constraints or averaged null energy conditions and therefore these theories are non-unitary. Similar statements can also be found for conformal higher derivative fields \cite{Mukherjee:2021alj}.
\section{Quasinormal modes and bulk thermodynamics}
In this section, we study the quasinormal spectrum of conformal higher spins in the static $dS$ patch.  For any inertial observer, the $dS$ vacuum appears to be thermal \cite{Denef:2009kn} and the partition function or the particularly the bulk-character encodes the spectrum of the quasinormal modes \cite{Anninos:2020hfj}. In \cite{Denef:2009kn} it was shown how to obtain the partition function from the quasinormal modes. Here one asks the other way which is given the character integral representation of the partition function how one extracts the quasinormal modes. 

From the  expansion of the bulk character $\chi^{b,\rm{dS}}_{s,\rm{Weyl}}$ one can read off the spectrum of quasinormal modes \cite{Anninos:2020hfj}
\begin{align}
    \chi^{b,\rm{dS}}_{s,\rm{Weyl}}&=\sum_r N_r x^r,
\end{align}
where $N_r$ is the degeneracy of the quasinormal modes decaying as $e^{-|t|r}$.
One can also extract the spectrum of the quasinormal modes from the resonance poles of the Fourier transform of the bulk character \cite{Anninos:2020hfj}.
\begin{align}
  \rho(\omega)=\frac{1}{2\pi}\int_{-\infty}^{\infty} dt \chi^{b,\rm{dS}}_{s,\rm{Weyl}}e^{i\omega t}.
\end{align}
The normal mode density $\rho(\omega)$ has poles at $\omega=\pm ir$ and quasinormal modes have resonance at these poles. Therefore one can obtain the spectrum of the quasinormal modes from the bulk character. 

In the previous section, we have observed that the bulk partition of Weyl graviton and conformal higher spins on $S^4$ agrees with the partition function on the hyperbolic cylinders. Therefore we take the character of the Weyl graviton on the hyperbolic cylinder which is the same as the bulk character on $S^4$ and expand it around $x=0$ to obtain the quasinormal mode spectrum. But before that, we `flip' the character using the procedure given in \cite{Anninos:2020hfj}.
\begin{align}
    [\chi]^{b,\rm{dS}}_{+,2,\rm{Weyl}}&=\chi^{b,\rm{dS}}_{2,\rm{Weyl}}-c_0-c_{\ell}(x^{\ell}+x^{-\ell})\nonumber\\
    &=\frac{2 x^2 \left(-4 x^2+5 x+5\right)}{(1-x)^3}\nonumber\\
    &=10x^2+40 x^3+82 x^4+136 x^5+202 x^6+280 x^7+370 x^8+\cdots
\end{align}
We read off the quasinormal frequencies of Weyl graviton $\omega_{\rm{Weylgrav}}$ in the static $dS$ patch in four dimension and the degeneracy $N_r$ from the expansion. The spectrum is given by
\begin{align}
    \omega_{\rm{Weylgrav}}=\pm2i, \pm3i, \pm4i,\pm5i,\cdots
\end{align}
Now to compare the spectrum of the quasinormal frequencies of the massless graviton, we take the `flipped' bulk character of massless graviton on $S^4$ given in \cite{Anninos:2020hfj} and expand it around $x=0$  to obtain
\begin{align}
     [\chi]^{b,\rm{dS}}_{+,2}&=\frac{2 x^3 (-3 x+5)}{(1-x)^3}\nonumber\\
     &=10 x^3+24 x^4+42 x^5+64 x^6+90 x^7+120 x^8+\cdots
\end{align}
Therefore the quasinormal frequencies of the massless graviton in the $dS_4$ static patch are given by
\begin{align}
    \omega_{\rm{grav}}&=\pm3i, \pm4i,\pm5i,\cdots
\end{align}
Note that the quasinormal frequencies of the massless graviton in static $dS_4$ patch start from $\pm 3i$ and it contains all the imaginary integers but misses out the $\pm 2i$ mode which is present in the spectrum of Weyl graviton. Similarly, one can also extract the spectrum of quasinormal modes of conformal higher spins in a static patch of $dS_4$ from the expansion of the bulk character given in \eqref{chsch}. We evaluate the spectrum  of quasinormal modes for conformal spin-$3$ and conformal spin-$4$ field explicitly. We evaluate the `flipped' bulk character for conformal spin-$3$ and expand it around $x=0$
\begin{align}
    [\chi]^{b,\rm{dS}}_{+,3,\rm{Weyl}}&=\chi^{b,\rm{dS}}_{3,\rm{Weyl}}-c_0-c_{\ell}(x^{\ell}+x^{-\ell})\nonumber\\
    &=\frac{2 x^2 \left(9 x^3-7 x^2-7 x-7\right)}{(x-1)^3}\nonumber\\
    &=14 x^2 + 56 x^3 + 140 x^4 + 248 x^5 + 380 x^6 + 536 x^7 + 716 x^8+\cdots
\end{align}
We extract the quasinormal spectrum of conformal spin-$3$
\begin{align}
    \omega_{3,\rm{Weyl}}=\pm2i, \pm3i, \pm4i,\pm5i,\cdots
\end{align}
Similarly, we also extract the quasinormal spectrum of conformla spin-$4$ field which turns out to be the same of the Weyl graviton and conformal spin-$3$
\begin{align}
    \omega_{4,\rm{Weyl}}=\pm2i, \pm3i, \pm4i,\pm5i,\cdots
\end{align}
We extract the quasinormal frequencies of all conformal spin fields and observe that the spectrum starts from $\pm 2i$. 
To compare it with the spectrum of the quasinormal modes of massless higher spins in the $dS_4$ static patch, we evaluate the `flipped' character of conformal higher spin fields and expand it around $x=0$. One can  observe that the  spectrum of massless higher spin-$s$ field starts from $\pm i(s+1)$. 
\begin{align}
    [\chi]^{b,\rm{dS}}_{+,s}&=2 \frac{(2 s+1) x^{s+1}-(2 s-1) x^{s+2}}{(1-x)^3}\nonumber\\
    &=2 x^{(1 + s)} + 4 s x^{(1 + s)} + 8 x^{(2 + s)} + 8 s x^{(2 + s)} + 
 18 x^{(3 + s)} + 12 s x^{(3 + s)}+\cdots
\end{align}
Therefore the quasinormal frequencies of massless higher spin fields are
\begin{align}
    \omega_s&=\pm i(s+1), \pm i(s+2), \pm i (s+3)+\cdots
\end{align}

Quasinormal spectrum is also useful to construct the characters of the corresponding theories.
It was shown that the characters obtained from the quasinormal spectrum of massless higher spins are identical to that of Harish-Chandra character of unitary massless spin-$s$ field \cite{Sun:2020sgn}. But we observe that the quasinormal spectrum of conformal higher spin fields contains $s-1$ extra modes compared to the spectrum of the unitary massless higher spin fields. Therefore one concludes that conformal higher spins in static $dS_4$ patch contain $s-1$ number of extra distinct states compared to the unitary massless higher spin fields.

One can evaluate the bulk thermodynamic quantities from the spectrum of the quasinormal modes but it will be easier to obtain those quantities from the logarithmic divergent part of the bulk free energy of Weyl graviton and conformal higher spins on $S^4$. Since we have observed the characters are identical on hyperbolic cylinders, we work with the partition function on the hyperbolic cylinders.

The logarithmic correction of the bulk entropy at horizon equilibrium can be obtained as
\begin{align}
    S_{\rm{bulk}}&=(1-q\partial_q)\log\mathcal{Z}^{b,\rm{dS}}_{s,\rm{Weyl}}(\rm{bulk)}|_{q=1}\nonumber\\
    &=-\frac{1}{90} s (s+1) (5 s (s+1)+6).
\end{align}
Here $\beta=2\pi q$ is the inverse temperature. Note that this is the same as the entanglement entropy of conformal higher spins in $d=4$ dimension given in \eqref{eechs}. Therefore one can interpret logarithmic correction of the bulk de Sitter horizon entropy as the universal contribution of the entanglement entropy of the conformal higher spin fields.
The universal part of the bulk energy at the horizon equilibrium can be obtained
\begin{align}
    U^{\rm{univ}}_{\rm{bulk}}&=-\frac{1}{2\pi}\partial_q\log\mathcal{Z}^{b,\rm{dS}}_{s,\rm{Weyl}}(\rm{bulk)}|_{q=1}\nonumber\\
    &=\frac{(s-1) s (s+1) (s+2) (3 s (s+1)+2)}{120 \pi }.
\end{align}
From the bulk entropy, we evaluate the bulk heat capacity at the horizon equilibrium
\begin{align}
    C_{\rm{bulk}}&=-q^2\partial_q^2\log\mathcal{Z}^{b,\rm{dS}}_{s,\rm{Weyl}}(\rm{bulk)}|_{q=1}\nonumber\\
    &=\frac{1}{90} s (s+1) (5 s (s+1)+8).
\end{align}

Note that all the bulk thermodynamic quantities extracted from the bulk character of $S^4$ is  the same if one uses the free energy of the hyperbolic cylinder. 

\section{Conclusions}
We find the relationship of the partition functions of the conformal higher spins on the spaces which are conformally related in $d=4$ dimension. We find the character integral representation of the partition function is a convenient way to compare them. We observe that the partition function of conformal higher spins on $S^4$ consists of bulk and edge characters. However, the partition function of conformal higher spins on the hyperbolic cylinder captures only the bulk character of the partition function. 

This observation is important in the context of evaluation of the entanglement entropy of conformal fields across a spherical surface. For the free Maxwell field in $d=4$ dimension, one can obtain the extractable part of the entanglement entropy \cite{Soni:2016ogt} which coincides with that obtained from the free energy on the hyperbolic cylinder \cite{ Huang:2014pfa, David:2020mls}. Therefore extractable part of the entanglement entropy comes from the bulk partition function  \cite{David:2021wrw} on the sphere. So one can interpret the bulk entropy of de Sitter space at horizon equilibrium as the entanglement entropy \cite{Anninos:2020hfj}. This property seems general even for the Weyl graviton and conformal higher spins. It will be interesting to show that the non-extractable part of the entanglement entropy coincides with the edge character of conformal higher spins on the sphere in an arbitrary dimension.

Using the conformal dimension of the twist operator of the conformal higher spins, we show that these theories do not obey the averaged null energy condition. We obtain a linear relation of Hofman-Maldacena variables which represents a straight line on the plane. The conformal higher spin theory is a point on the straight line. We observe that the straight line never intersects the region of unitarity which is a triangular region in the plane. Therefore, the conformal higher spin field can not be inside this region. From the bulk character of the conformal higher spins on $S^4$, we extract the spectrum of quasinormal modes. We observe that this spectrum consists of $s-1$ number of extra distinct states compared to the unitary massless higher spin fields on $S^4$. We also evaluate the bulk thermodynamic entropy of conformal higher spins which turns out to be the same as the entanglement entropy across a spherical surface.
\appendix
\section{Gauge fixing of Weyl graviton on $S^4$}\label{gaugefixs}
It is convenient to  first decompose the field in pure trace and traceless part
\begin{eqnarray}\label{defth4}
h_{\mu\nu}  &=& \bar h_{\mu\nu} + \frac{1}{4} g_{\mu\nu} h,  \\ \nonumber
\bar h_{\mu\nu}  &=& h_{\mu\nu}^\perp + \nabla_{\mu} \zeta^\perp_{\nu} + \nabla_{\nu}\zeta^\perp_{\mu}  +
 \nabla_{\mu} \nabla_{\nu} \sigma -\frac{1}{4} g_{\mu\nu} \Box\sigma ,  \\ \nonumber
 & &{\rm where} \qquad  \nabla^{\mu} h_{\mu\nu}^\perp = 0, \qquad \nabla^{\mu} \zeta^\perp_{\mu} =0.
\end{eqnarray}
 The traceless part $\bar h_{\mu\nu}$ is further decomposed into transverse and longitudinal parts. Using these decomposed field variables the transverse part of the action takes the form
\begin{align}
    \mathcal{L}^{(2)}&=h_{\mu\nu}^\perp \mathcal{O}_2h_{\mu\nu}^\perp,
\end{align}
where $\mathcal{O}_2$ is the fourth order operator given in \eqref{4op}.\\
The change of variables involves a Jacobian. We start with the canonical definition of he measure given by the integral
\begin{align}\label{gravnorm}
\int   {\cal D} h_{\mu\nu} e^{-\int d^4x{\sqrt{g}} (   h_{\mu\nu} h^{\mu\nu}) 
}=1.
\end{align}
Now we define $\tilde{h}=\Box\sigma-h$ and therefore
\begin{align}
    h_{\mu\nu}=h_{\mu\nu}^\perp + \nabla_{\mu} \zeta^\perp_{\nu} + \nabla_{\nu}\zeta^\perp_{\mu}  +
 \nabla_{\mu} \nabla_{\nu} \sigma -\frac{1}{4} g_{\mu\nu} \tilde{h}
\end{align}
 With these definitions we can expand the integral
 \begin{eqnarray}
& &  \int d^4 x \sqrt{g} h_{\mu\nu} h^{\mu\nu} =  \int d^4 x \sqrt{g} \left[  h_{\mu\nu}^\perp h^{\perp\, \mu\nu}  + 
 (\nabla_{\mu}\zeta^{\perp}_{ \nu}   +\nabla_{\nu}\zeta^{\perp}_{\mu})(\nabla^{\mu}\zeta^{\perp \nu}   +\nabla^{\nu}\zeta^{\perp\mu})  \right. \\ \nonumber
 & & \qquad\qquad\qquad\qquad \left. 
 + (\nabla_{\mu}\nabla_{\nu}\sigma)(\nabla^{\mu}\nabla^{\nu}\sigma)+\frac{1}{4}\tilde{h}^2\right] 
 \end{eqnarray}
 The cross terms cancel due to integration by parts imposing the transverse gauge and traceless conditions.
 Substituting this 
 we can further simplify the terms by integration by parts and obtain
 \begin{eqnarray}\label{expansgrav}
 \int d^4 x \sqrt{g} h_{\mu\nu} h^{\mu\nu} &=&  \int d^4 x \sqrt{g} \left[  h_{\mu\nu}^\perp h^{\perp\, \mu\nu} 
 + 2 (\zeta_{\mu}^{~\perp} (-\Delta_{(1)} )  \zeta^{i\perp}+3\zeta^{i\perp}\zeta_{i}^{~\perp})
 \right. \\ \nonumber
 && +\sigma \Delta_{(0)}(\Delta_{(0)}-4)\sigma+\frac{1}{4}\tilde{h}^2.
 \end{eqnarray}
 Note that, we have  redefined the variable $\tilde{h}=\Box\sigma-h$. The Jacobian associated with this is the massless scalar Laplacian which will get cancelled with the quadratic part of sigma in \eqref{expansgrav}. Now combining all the Jacobians from the change in variables, the partition function becomes
 \begin{align}
    \mathcal{Z}_{2,\rm{Weyl}}^{(S^4)}&=\left(\frac{\det\Delta_{(1)}^{\perp}(-3)\det\Delta_{(0)}(-4)}{\det\Delta_{(2)}^{\perp}(4)\det\Delta_{(2)}^{\perp}(2)}\right)^{\frac{1}{2}}  .
 \end{align}
 Note that, transverse spin-1 determinant comes from the integral over $\zeta^{\perp \mu}$ in the change in measure and spin-0 determinant comes from the integration over the $\sigma$ variable. The same analysis will go through for any Einstein space in $d=4$ dimension with the substitution of $R_{\mu\nu}=\frac{1}{4}g_{\mu\nu}R$. We have carried out for $AdS_4$ and obtained the partition function.
\section{Change in measure for Weyl graviton on $AdS_3\times S^1$}\label{appa}
We  evaluate the change in measure which comes due to the fact that we have decomposed the field variables into transverse and longitudinal parts and therefore, change in variables will involve a Jacobian.   

The canonical definition of the measure fixes the normalization
\begin{align}\label{gravnorm}
\int \mathcal{D} h_{\mu\nu} e^{-\int\sqrt{g} d^4 x h_{\mu\nu} h^{\mu\nu}}=1.
\end{align}
We first decompose $h_{\mu\nu}$ along the $S^1$ as well as $AdS_3$ direction.  We denote ${i,j}$ along the $AdS_3$ direction.
 \begin{align}
 h_{\mu\nu}h^{\mu\nu}=(h_{00})^2+(h_{0 i})^2+h_{ij}h^{ij}.
 \end{align}
We write $h_{ij}$ as a pure trace and traceless part and the pure trace is again decomposed in transverse and longitudinal parts as given in \eqref{defth}.
Therefore the integrand in the exponential of \eqref{gravnorm} can be written as
 \begin{align}\label{expansgrav3}
  h_{\mu\nu}h^{\mu\nu}&=(h_{00})^2+(h_{0 i})^2+(h^{\perp}_{~ij})^2+(\nabla^{i}\zeta^{\perp j}+\nabla^{j}\zeta^{\perp i})^2
  \nonumber\\
                      &+(\nabla^{i}\nabla^{j}\sigma-\frac{1}{3}g^{ij}\Box\sigma)^2+\frac{1}{3}h^2\nonumber\\
                      &=(h_{00})^2+(h_{0 i})^2+(h^{\perp}_{~ij})^2+2 (-\zeta_{i}^{~\perp}\Delta_1 \zeta^{i\perp}+2\zeta^{i\perp}\zeta_{i}^{~\perp})\nonumber\\
                      &+\frac{2}{3}\sigma \Delta_{0}(\Delta_{0}-3)\sigma+\frac{1}{3}h^2.
 \end{align}
 Here $\Delta_{0}$ and $\Delta_{1}$ scalar and spin-$1$ transverse Laplacian on $AdS_3$.
  Note that all the cross terms vanish by integration by parts and imposing transverse gauge condition. Now we integrate over all the independent variables $h_{00}$,$h_{0i}$, $h^{\perp}_{~ij}$, $\zeta_{i}^{~\perp}$, $\sigma$ and $h$.  We plug (\ref{expansgrav3}) in (\ref{gravnorm}) and get change in measure as
$$\lvert J\rvert=\Big((-\Delta_{1}+2)\Delta_{0}(\Delta_0-3)\Big)^{\frac{1}{2}}.$$

 \acknowledgments
 The author  thanks Justin R. David for  discussions, useful comments on the manuscript and encouragement. He also thanks Debajyoti Sarkar for the discussion.
\bibliographystyle{JHEP}
\bibliography{CHS.bib}
\end{document}